\documentclass{article}
\usepackage[english]{babel}
\usepackage[a4paper,top=2cm,bottom=2cm,left=3cm,right=3cm,marginparwidth=1.75cm]{geometry}

\usepackage{amsmath}
\usepackage{graphicx}
\usepackage[colorlinks=true, allcolors=blue]{hyperref}
\usepackage[affil-it,blocks]{authblk}
\usepackage{siunitx}
\usepackage{ulem}
\usepackage{ mathrsfs }
\usepackage{float}
\usepackage[caption = false]{subfig}
\usepackage{pdfpages}

\usepackage[sorting=none]{biblatex} 
\usepackage[mathlines]{lineno}

\usepackage{orcidlink}
\usepackage{placeins}

\renewcommand{\Re}{\operatorname{Re}}
\renewcommand{\Im}{\operatorname{Im}}

\title{Proposal for PAC 52:\\ Measurement of $\alpha_-$ for $\Lambda\rightarrow p\pi^-$}

\author{P.~Hurck\thanks{Contact: \href{mailto:Peter.Hurck@glasgow.ac.uk}{Peter.Hurck@glasgow.ac.uk}} \orcidlink{0000-0002-8473-1470}}
\author{D.I.~Glazier}
\author{D.G.~Ireland}
\author{K.~Livingston}
\affil{School of Physics and Astronomy, University of Glasgow, United Kingdom}

\author{F.~Afzal}
\author{A.~Thiel\orcidlink{0000-0003-0753-696X}}
\author{Y.~Wunderlich}
\affil{HISKP, University of Bonn, Germany}

\author{V.~Crede}
\affil{Florida State University, Tallahassee, Florida 32306, USA}

\author{M. M.~Dalton\,\orcidlink{0000-0001-9204-7559}}
\affil{Jefferson Lab, Newport News, USA}

\begin{document}
\maketitle

\begin{abstract}
We propose to measure the weak decay constant $\alpha_-$ for the decay $\Lambda\rightarrow p\pi^-$ using a both circularly and linearly polarized photon beam with the GlueX spectrometer in Hall D. 
The measurement will take advantage of the fact that a measurement with both linear and circular photon beam polarization results in an over-constrained set of amplitudes which can be fitted to data and used to extract $\alpha_-$ which will be left as a free parameter in the fit.
We expect to determine $\alpha_-$ with statistical uncertainties comparable to existing measurements and independent systematic uncertainties.
This measurement can be performed alongside GlueX\nobreakdash-II running and requires no new hardware or new beam time.  
The measurement requires that a sufficient fraction of the electron beam polarization be longitudinal in the Hall D tagger.

\end{abstract}


\tableofcontents

\section{Introduction}
The decay parameter $\alpha_-$ of the parity-violating weak decay $\Lambda\to p\pi^-$ describes the interference between parity-violating $s$~and parity-conserving $p$~waves. Among other things, the parameter $\alpha_-$ of the singly-strange $\Lambda$~hyperon is an important quantity for the extraction of polarization observables in various experiments. Many other hyperons exhibit a $\Lambda$ in their decay chain, e.g. in the prominent decays $\Sigma\to\Lambda\gamma$, $\Xi^{0(-)}\to\Lambda\pi^{0(-)}$, and $\Omega\to\Lambda K^-$, and therefore, the decay parameters of these hyperons are strongly affected by $\alpha_-$. In general, the parameter $\alpha_-$ affects any quantity in which the polarization of the $\Lambda$ is relevant. For this reason, an independent determination of this quantity is highly desirable given that $\alpha_-$ plays an important role in various fields of physics. For instance, comparing $\alpha_-$ with the parameter $\alpha_+$, which originates from the charge-conjugate decay $\bar{\Lambda} \to\bar{p}\pi^+$, provides a test of CP symmetry for strange baryons and, thus, can potentially shed light on the matter-antimatter asymmetry in the Universe~\cite{Sakharov:1967dj}.

Small violations of CP symmetry are predicted by the standard model and are a well established phenomenon in weak decays of mesons. However, the mechanisms of the standard model are too specific to yield effects of a size that can explain the observed matter–antimatter asymmetry of the Universe. Therefore, CP tests can be considered a promising area to search for physics beyond the standard model. And so far, no CP-violating effects beyond the standard model have been observed in the baryon sector. In this respect, a CP violation at the $3.3\sigma$ level has been found by the LHCb Collaboration in four-body decays of $\Lambda^0_b$ and $\bar{\Lambda}^0_b$~baryons~\cite{LHCb:2016yco}. However, in the BESIII simultaneous measurement of $\alpha_-$ and $\alpha_+$ of the $\Lambda$, no sign of any CP violation was found~\cite{BESIII:2018cnd}.

\begin{figure}[h]
    \centering
    \includegraphics[width=0.6\linewidth]{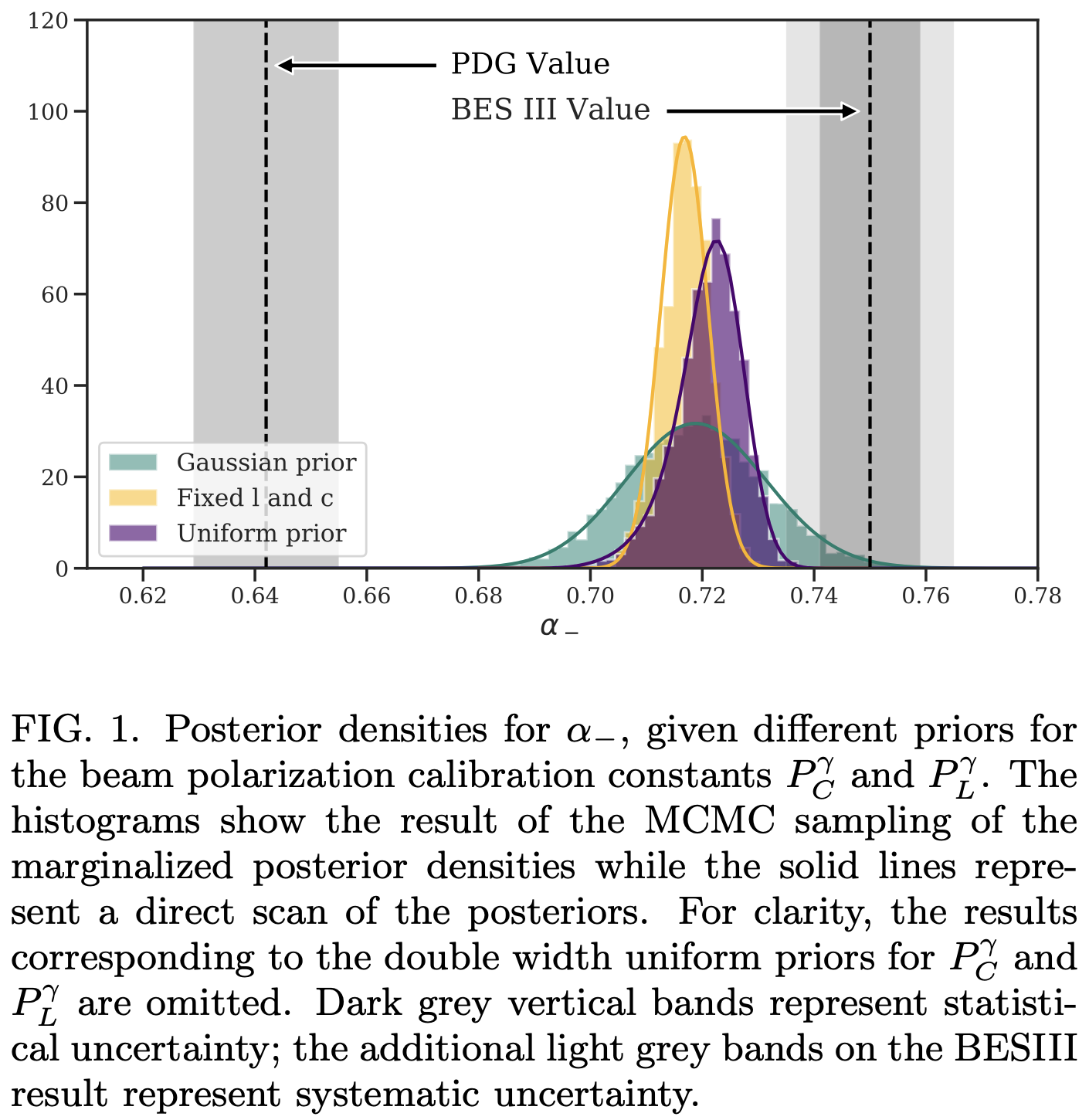}
    \caption{Discrepancy between old and new measurements of $\alpha_-$. The refit results based on $\Lambda$ photoproduction at CLAS6 are shown as posterior densities. Taken from \cite{Ireland:2019uja} (including caption).}
    \label{fig:alpha_world_data}
\end{figure}

Our goal is to measure the weak decay constant $\alpha_-$ for $\Lambda\rightarrow p\pi^-$, using the photoproduction reaction $\gamma p \to K^+\Lambda$. The measurement will be carried out with an elliptically polarized photon beam with a linear and circular polarization component and  using the GlueX spectrometer in Hall~D. Having both linear and circular photon beam polarization results in an over-constrained set of amplitudes which can be fitted to the $K^+\Lambda$ data and used to extract $\alpha_-$, which will be left as a free parameter in the fit. 

We expect that the gathering of this data will have no adverse impacts on the overall running of GlueX\nobreakdash-II and we do not expect to require any additional dedicated time for systematic studies.

The proposal is structured as follows.  
In Sec.~\ref{sec:existing} we survey the landscape of existing data on $\alpha_{-}$.  
In Sec.~\ref{sec:method} we detail the method for determining $\alpha_{-}$, which has independent systematic uncertainties from the existing measurements with highest precision.
In Sec.~\ref{sec:beam} we describe the requirements for beam circular polarization and polarimetery, which is the only difference from the approved GlueX\nobreakdash-II running.
In Sec.~\ref{sec:PoC} we use a small subset of existing data taken in 2023 to determine a value of $\alpha_-$, demonstrating the feasibility and obtaining a statistical uncertainty to use in projections.  
We use the same data set to determine an asymmetry in $\gamma p\to\rho p$ proportional to the circular polarization.
In Sec.~\ref{sec:experiment} we do a projection of the statistical uncertainty we would obtain from the existing data and the remainder of the GlueX\nobreakdash-II running if we receive an electron beam polarization of 80\%.  The projected statistical uncertainty  is comparable to the uncertainties on existing measurements.
In Sec.~\ref{sec:additionalphysics} we present some additional physics opportunities that would become possible with a circularly polarized photon beam as requested in this proposal.

\section{Existing measurements}
\label{sec:existing}
The $\alpha_-$~parameter was first measured in the 1960s at Brookhaven in the reaction $\pi^- p \to \Lambda K^0$ using a carbon-plate spark chamber. The value of $\alpha_- = 0.62\pm 0.07$ was extracted from observing an up-down asymmetry of the $\Lambda$~decay pions~\cite{Cronin:1963zb}, which demonstrated that the reaction produced polarized $\Lambda$~hyperons. Subsequent experiments in the late 1960s at the Princeton-Pennsylvania Accelerator~\cite{Overseth:1967zz} using the same reaction and at the Lawrence Radiation Laboratory~\cite{Dauber:1969hg} using the decay of the $\Xi$~hyperon into $\Lambda\pi$ provided larger $\Lambda\to p\pi^-$~event samples. The determined values of $0.645\pm 0.017$ and $0.67\pm 0.06$, respectively, were in good agreement with the earlier measurement. Two additional measurements in the 1970s at CERN of $0.649\pm 0.023$~\cite{Cleland:1972fa} and $0.584\pm 0.046$~\cite{Astbury:1975hn} provided further results for $\alpha_-$, again fairly consistent with the known values at the time. 

More recently, the BESIII Collaboration in 2019 reported a significantly larger value of $0.750\pm 0.009\pm 0.004$~\cite{BESIII:2018cnd}, which was inconsistent with the Review of Particle Physics (RPP) average value of $0.642\pm 0.013$ quoted until 2018~\cite{ParticleDataGroup:2018ovx}. Since these two values had uncertainties at the percent level, the $5\sigma$~discrepancy rendered these results incompatible. The about 17\,\% higher value reported by BESIII in 2019 has triggered a whole new series of measurements at various laboratories. In the same year, a study by Ireland {\it et al.}~\cite{Ireland:2019uja} based on a sample of photoproduced $K^+\Lambda$~events off the proton collected by the CLAS Collaboration at Jefferson Lab supported a higher value, but is lower than and in tension with the BESIII value. The obtained value of $0.721(6)(5)$ was corroborated by multiple statistical tests as well as a modern phenomenological model, showing that the new value yielded the best description of the data in question. In 2020, the LHCb Collaboration presented an analysis of the $\Lambda_b^0\to J/\psi \Lambda$ angular distribution and the transverse production polarization of $\Lambda_b^0$~baryons in proton-proton collisions at centre-of-mass energies of 7, 8 and 13~TeV~\cite{LHCb:2020iux}. The parity-violating asymmetry parameter of the $\Lambda\to p\pi^-$ decay was also determined from the same data and its value of $0.74^{+0.04}_{-0.03}$ found to be consistent again with the recent 2019 measurement by the BESIII collaboration. Finally, the 2022 report by the BESIII Collaboration presented the most precise measurements of $\Lambda$~decay parameters and CP asymmetry with a five-dimensional fit to the full angular distributions of the daughter baryon~\cite{BESIII:2022qax}. The extracted $\alpha_-$~value of $0.7519\pm 0.0036\pm 0.0024$ has an unprecedented statistical quality and the smallest systematic uncertainty reported to date. The decay parameter $\alpha_+$ was also reported. The results are based on a total sample of 10 billion $J/\psi$~events and about 3.2 million quantum-entangled $\Lambda$-$\bar{\Lambda}$ pairs could be fully reconstructed in the decay $J/\psi\to\Lambda\bar{\Lambda}$ with $\Lambda(\bar{\Lambda})\to p\pi^- (\bar{p}\pi^+)$. The most recent analysis by the BESIII provides further support for the previous BESIII results. A value of $0.757\pm 0.011\pm 0.008$ based on $J/\psi \to\Xi\bar{\Xi}\to \Lambda\bar{\Lambda}\pi\pi$ was reported in late 2022~\cite{BESIII:2021ypr}. 

As discussed in Ref.~\cite{Ireland:2019uja}, the big discrepancy between the earlier RPP~value based on measurements in the 1960s and 70s, and the more recent results might be due, for instance, to underestimated systematic effects in the calculation of correction factors in Ref.~\cite{Astbury:1975hn}. Or in the case of Ref.~\cite{Cleland:1972fa}, photographs of carbon-plate spark chambers were used, and a ten-parameter kinematic fit applied to each event; several sources of uncertainty were highlighted and together with the approximate fitting method, there was ample scope for systematic error. 
While the previous measurements were all state of the art when carried out, the RPP 2023 online update lists only some of the more recent measurements “above the line.”

In a brief summary, several recent measurements of the $\alpha_-$~parameter have addressed the large discrepancy between the old results from the 1960s and 1970s, and a significantly larger value (by about 15--17\,\%), first observed by the BESIII Collaboration in 2019, in support of the larger value. However, the experimental status of $\alpha_-$ is not yet satisfactorily resolved. The Particle Data Group has not used the 2020 LHCb result~\cite{LHCb:2020iux} for their average, likely due to the reported fairly large uncertainties. For this reason, the quoted RPP average is only based on various BESIII measurements and the result by Ireland {\it et al.} based on CLAS data~\cite{Ireland:2019uja}. While the reported BESIII results are all in good agreement and self-consistent, a smaller, but significant, discrepancy still persists. Ireland {\it et al.} have reported a value that is about 4\,\% lower than the averaged BESIII value. To this effect, it is worth noting that the methodology used by Ireland {\it et al.} was different. GlueX is well positioned to address this remaining discrepancy by providing an independent cross-check of the methodology discussed in Ref.~\cite{Ireland:2019uja}.

\section{Extraction method} 
\label{sec:method}

The methodology used for the extraction of $\alpha_-$ is based on a publication by D.~Ireland et al \cite{Ireland:2019uja}. Unfortunately, the literature is plagued by a variety of different sign conventions for polarization observables for single-pseudoscalar production. Therefore, we lay out the whole formalism with all chosen conventions in detail in the following.\par

The weak decay parameter $\alpha_-$ can be extracted by fitting polarization observables, using an elliptically polarized photon beam. The differential cross-section for single-pseudoscalar photoproduction using a beam with circular and linear polarization is given by~\cite{Barker:1975bp}
\begin{align}
    \rho_f \frac{\text{d}\sigma}{\text{d}t} = \left.\frac{\text{d}\sigma}{\text{d}t}\right\vert_\text{unpolarized}\{1 &+ \sigma_y P - P^\gamma_L\cos(2\Phi)(\Sigma+\sigma_y T) \nonumber \\
    & -P^\gamma_L\sin(2\Phi)(O_x\sigma_x + O_z\sigma_z) - P^\gamma_C(C_x\sigma_x + C_z\sigma_z)\}
\end{align}
where $P^\gamma_L$ is the transverse polarization of the beam at an angle $\Phi$ to the reaction plane and $P^\gamma_C$ is the degree of right circular polarization of the beam. The density matrix of the recoiling particle is denoted as $\rho_f = \frac{1}{2}(I + \sigma\cdot P_f)$ and $P_f$ is its polarization. \par
In order to fit the function using an event-based likelihood, we define the intensity function
\begin{align}
    I(\Phi,\theta_{x',y',z'}) &= 1+\textcolor{cyan}{\alpha_-} \cos\theta_{y'} \textcolor{blue}{P} - P^\gamma_L\cos(2\Phi)(\textcolor{blue}{\Sigma}+\textcolor{cyan}{\alpha_-}\cos\theta_{y'} \textcolor{blue}{T}) \nonumber \\
    & \hspace{1.5cm}-P^\gamma_L\sin(2\Phi)(\textcolor{cyan}{\alpha_-}\cos\theta_{x'} \textcolor{blue}{O_{x'}}+\textcolor{cyan}{\alpha_-}\cos\theta_{z'} \textcolor{blue}{O_{z'}}) \nonumber \\
    & \hspace{1.5cm}-P^\gamma_C(\textcolor{cyan}{\alpha_-}\cos\theta_{x'} \textcolor{blue}{C_{x'}}+\textcolor{cyan}{\alpha_-}\cos\theta_{z'} \textcolor{blue}{C_{z'}}) \label{eq:intensity}
\end{align}

This expression contains seven polarization observables $O_j\in\{\Sigma, T, P, C_{x'}, C_{z'}, O_{x'}, O_{z'}\}$, which depend on the angles $\Phi$ and $\theta_{x',y',z'}$ (c.f. Sec.~\ref{sec:proofofconcept}), as well as $P_L^\gamma$ and $P^\gamma_C$. The expression also contains the weak decay parameter $\alpha_-$, which we want to determine. Measuring $\frac{\text{d}\sigma}{\text{d}t}$ allows to extract all $O_j$, provided $\alpha_-$ is known. If it is not known, it can be left as a free parameter in the fit, but this causes Eq.~\eqref{eq:intensity} to be under-constrained. To remedy the situation two so-called Fierz identities which place constraints on the polarization observables can be exploited~\cite{Sandorfi:2010uv}:
\begin{align}
    \Sigma^2 - T^2 + P^2 + C_{x'}^2 + C_{z'}^2 + O_{x'}^2 + O_{z'}^2 &= 1 \label{eq:fierz1}\\
    \Sigma P - T - C_{x'}O_{z'} + C_{z'}O_{x'} &= 0 \label{eq:fierz2}
\end{align}
Using Eq.~\eqref{eq:intensity} and imposing the relations in Eqs.~\eqref{eq:fierz1} and \eqref{eq:fierz2} allows us to determine $\alpha_-$ from a fit to the data.\par

While writing the cross-section and intensity in terms of polarization amplitudes nicely illustrates the methodology, it is more convenient to directly fit the underlying transversity amplitudes. These do automatically contain the Fierz identities and hence they provide a more straightforward way to analyze the data without constraining the fits explicitly. They are given by
\begin{align}
    \frac{\text{d}\sigma}{\text{d}t} &= |b_1|^2 + |b_2|^2 + |b_3|^2 + |b_4|^2 \\
    \Sigma \frac{\text{d}\sigma}{\text{d}t} &= |b_1|^2 + |b_2|^2 - |b_3|^2 - |b_4|^2\\
    T \frac{\text{d}\sigma}{\text{d}t} &= |b_1|^2 - |b_2|^2 - |b_3|^2 + |b_4|^2\\
    P \frac{\text{d}\sigma}{\text{d}t} &= |b_1|^2 - |b_2|^2 + |b_3|^2 - |b_4|^2\\
    O_{x'} \frac{\text{d}\sigma}{\text{d}t} &= -2\Re(b_1b_4^* - b_2b_3^*)\\
    O_{z'} \frac{\text{d}\sigma}{\text{d}t} &= -2\Im(b_1b_4^* + b_2b_3^*)\\
    C_{x'} \frac{\text{d}\sigma}{\text{d}t} &=  2\Im(b_1b_4^* - b_2b_3^*)\\
    C_{z'} \frac{\text{d}\sigma}{\text{d}t} &= -2\Re(b_1b_4^* + b_2b_3^*)\\
\end{align}
In practice, these equations are rearranged for the polarization observable and inserted into Eq.~\eqref{eq:intensity}. The resulting intensity can be used for likelihood fitting. While the amplitudes are under-constrained the resulting polarization observables are not.

\subsection{Advantage over previous measurement with similar methodology}
As discussed before, our methodology is based on the same idea as used by Ireland et al. However, they did not have access to actual event level data. Instead they used published result for the polarization observables $\Sigma$, $T$, $P$, $O_x$, $O_z$, $C_x$, and $C_z$. Of these, only the first five observables had a common region in $\{W, \cos\theta\}$ space, with 314 data points. In order to incorporate $C_x$ and $C_z$, a Gaussian Process was used to interpolate the available data. This interpolation made it possible to estimate $C_x$ and $C_z$ in the same $\{W, \cos\theta\}$ region as the other available data. This data was then used to extract $\alpha_-$ through the constraints provided by the Fierz identities. \par
Although a large amount of $K^+\Lambda$ events went into the measurement of the data used in their $\alpha_-$ estimation, the systematical uncertainties are potentially large and hard to assess. A reliable uncertainty estimation requires precise knowledge of all systematic uncertainties present in the original data, as well as their correlations. Since all used data points were reported by the CLAS experiment it cannot be ruled out, that they share dependent systematic errors. It is possible that this could account for the remaining discrepancy between the Ireland et al and the BESIII result. \par
Our methodology improves upon this by measuring all polarization observables in a single measurement. That means that we will be in control of and able to accurately estimate all systematic uncertainties. This will allow us to provide a well controlled measurement of $\alpha_-$ which is completely independent of the BESIII methodology.

\section{Photon beam}
\label{sec:beam}

This measurement requires the beam photons to have both linear and circular polarization, here referred to as elliptical polarization.
Elliptically polarized photons can be produced using longitudinally polarized electrons that are incident on a thin diamond radiator. 
The degree of linear and circular polarization of the beam photons will be determined to sufficient precision using existing hardware as described below.

\subsection{Measurement with elliptically polarized photons}

The amount of circular polarization that a beam photon carries depends on its energy according to~\cite{olsen}:
\begin{equation}
    \label{eq:circpol}
    P^\gamma_C = p_e(4x-x^2)/(4-4x+3x^2),
\end{equation}
where $x=E_\gamma/E_e$ is the ratio of beam photon $E_\gamma$ to electron beam energy $E_e$. Eq.~\eqref{eq:circpol} was derived for an amorphous radiator, when using a diamond radiator there is a small correction downwards~\cite{A2:2024ydg} (see Fig.~\ref{fig:circpol_dia} in Appendix~\ref{app:circpol}). 

The position of the coherent peak will be set for GlueX\nobreakdash-II running.  We will obtain a linear polarization of about 38\% in the beam region approximately 8.5--9.0~GeV depending on the final electron beam energy.  At this energy about 90\% of the electron polarization is transferred to the photon.

Recently, the A2 collaboration in Mainz successfully extracted polarization observables that require linearly or circularly polarized photons from the same data sample using an elliptically polarized photon beam \cite{A2:2024ydg}. 
We will use the same approach for the proposed measurement in order to extract all seven polarization observables given in Eq.~\eqref{eq:intensity} from a single data sample.

\subsection{Determination of Circular Polarization }
\label{sec:polarimetry_circ}

The degree of circular polarization follows directly from the electron beam polarization (with a small correction needed for the diamond lattice mostly in the energy region of the coherent peak, see Appendix~\ref{app:circpol}). We do not request any dedicated beamtime for systematic studies on the effect of the Coherent Bremsstrahlung on the degree of circular polarization as this will be achieved using amorphous running for a fraction of the data in line with past measurements.
We will determine the electron beam polarization using projections from polarization measurements in other halls and using the reaction $\gamma p\to\rho p$ in the main GlueX spectrometer.

As detailed in Secs.~\ref{sec:rhopolarimeter} and \ref{sec:PoC_rho}, this data on exclusive $\rho$ photoproduction will be obtained with high statistical precision and will be available in perpetuity for any period when physics data is available.  
A full analysis in terms of partial waves is capable of determining the linear and circular polarizations independently.
A much easier analysis of the helicity asymmetry is capable of producing a relative polarization, and if it is calibrated only once using some method, then it can be used for the whole experiment to provide the absolute polarization over all the running.

We anticipate calibrating this $\rho$ polarimeter by projecting the precession of the polarization measured in other halls (which is established to ~1\% percent accuracy in Halls A and C) through to Hall D.  The accuracy with which this can be done depends on the details of the running conditions.  Different energies in other halls, different linac energies, and particularly different Wien angles would all contribute to constraining the parameters.  

Studies are being done, for the future polarized target program in Hall D, into a potential future M\o ller polarimeter in the Hall D tagger---but this would not be necessary for this proposed measurement to achieve its goals.  Such a device allows an independent check of the other strategies outlined here and may benefit this program even if it came later.

\subsubsection{Projection from other measurements}

As described in more detail in Sec.~\ref{sec:PoC_data} it is possible to do a combined fit of polarization data obtained in other halls to fit the beam polarization at the source, the linac energies, and any Wien angle offset, which allows the precession to be projected to Hall D.
The accuracy that can be achieved using this approach depends on the details of what data is available for the fit and the procession angle in Hall D.
The more longitudinal the polarization in Hall D, the more accurate the projection will be.
Ideally, if multiple halls measure the polarization at multiple Wien angles and energies then there are sufficient independent constraints such that projection can be done with $<2\%$ uncertainty on the longitudinal polarization.

\subsubsection{Physics Reaction in the GlueX Spectrometer}
\label{sec:rhopolarimeter}

The circular polarization of the photons at the target can be measured in the main GlueX spectrometer using the reaction $\gamma p\to\rho p$.
This reaction has a very high statistics, corresponding to about 10\% of the total hadronic cross section in GlueX.
As described in Sec.~\ref{sec:PoC_rho}, the integrated helicity asymmetry can be determined with a statistical uncertainty of $dP^\gamma_{C}/P^\gamma_{C}=2.7\%$ per 2-hour run at 75\% polarization.  
This can be a very good relative polarimeter which allows us to monitor the polarization as a function of time.
The full resolution would only be available later, an online result from $\sim7\%$ of the data would  be available in real time giving an uncertainty limited to $dP^\gamma_{C}/P^\gamma_{C}=3.0\%$ per day for diagnostic purposes.

Sec.~\ref{sec:PoC_rho} also describes how a full analysis can be used to determine the beam polarizations. 
This requires well calibrated data and a well matched Monte Carlo simulation and hence will not be available online.
Systematic uncertainties in the degree of circular polarization can be controlled and minimized by comparing the $\rho$ measurements with polarization values obtained from other sources.
This can be done by doing a combined fit~\cite{Grames:2004mk} and leveraging the well understood systematic uncertainties of the M\o ller polarimeters in other halls to achieve a polarization systematic uncertainty in Hall D of $<2\%$.
Independently calibrating this ``$\rho$ polarimeter" is essentially equivalent to measuring a new spin-density matrix element (SDME) for $\gamma p\to\rho p$ reaction, a publishable result.

\subsection{Measurement of Absolute Linear Polarization }
\label{sec:polarimetry_linear}

The linear polarization of the beam is measured using the Triplet Polarimeter (TPol) which uses the process of $e^+e^-$ pair production on atomic electrons in a beryllium target foil.
On timescales of days this is a statistics limited measurement.  The estimated total systematic uncertainty is 1.5\%~\cite{Dugger:2017zoq} and over the course of a run a total uncertainty of 2.1\% has been achieved~\cite{GlueX:2017zoo}.

\section{Proof of concept}
\label{sec:PoC}

\subsection{Existing Data}
\label{sec:PoC_data}

During the running of GlueX\nobreakdash-II from January 12 to March 20 2023, the helicity signal was incorporated into the GlueX DAQ for the first time.  
This allows analysis that depends on the helicity of the electron beam to be performed.
During this time, 153 billion events were recorded taking  2,188\,TB of space (runs 120286 to 121207 of the 2023 beamtime in Hall D parlance.)

\begin{table}[ht]
    \centering
    \caption{Polarization versus run number for GlueX\nobreakdash-II in 2023.  The quoted uncertainties are very conservative pending more detailed studies of the precession and the $\rho$ polarimeter.  This is sufficient for the ``initial subset" of data that is analyzed here.}
    \begin{tabular}{c|c|c}
        runs &  Wien angle & longitudinal polarization \\ 
        \hline
        120286 to 120445 & -64.6 &  $53.2\pm4.0\%$ \\
        120446 to 121207 & -47.2 &  $71.0\pm4.0\%$ \\
    \end{tabular}
    \label{tab:polarization}
\end{table}

\begin{figure}[h]
    \centering
    \includegraphics[width=0.675\linewidth]{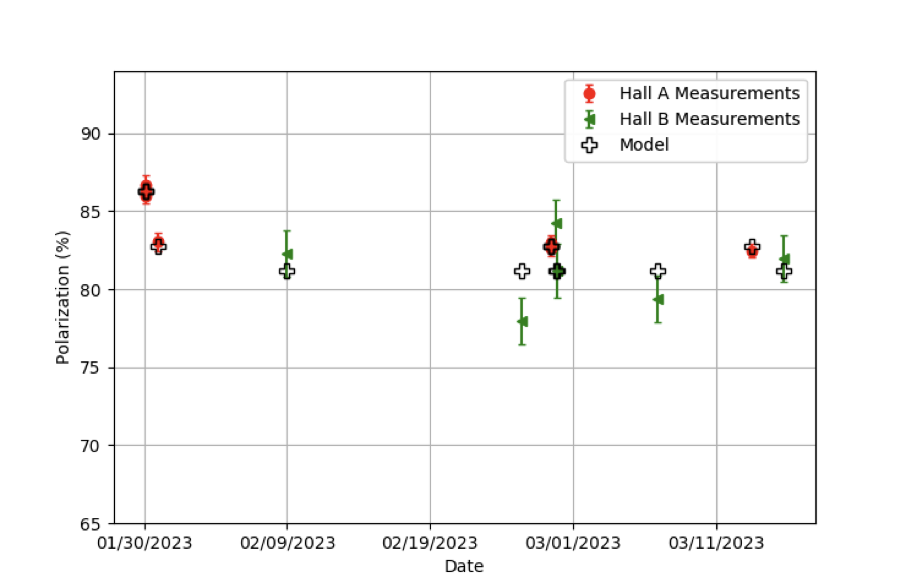}
    \caption{A fit of the M\o ller measurements made in early 2023 in Hall A and Hall B at 2 Wien angles to a global model of the beam energy and polarization magnitude and launch angle.}
    \label{fig:polfit}
\end{figure}

Fortuitously, the longitudinal polarization of the electron beam, when it arrived in the Hall~D tagger, was high, Table~\ref{tab:polarization}.
The polarization in Hall D was determined by doing a combined fit of the M\o ller polarimeter data in Hall A and Hall B for 2 Wien settings.
The fit uses the measured injector energy, the beam optic elements from a CEBAF Elegant model and takes into account synchrotron energy loss in the precession.
The parameters of the fit are the absolute beam polarization at the source, the linac energies, and the Wien angle offset.
The result of the fit can be used to be project the beam to Hall D, see Fig.~\ref{fig:polfit}

An ``initial subset" of the GlueX\nobreakdash-II 2023 data was chosen to do a preliminary analysis.  This subset is 30 good runs taken over 53 hours between 2023-01-27 to 2023-01-30 with numbers from 120395 to 120438, representing about 5\% of the data during that beam period.  This data was given an initial calibration during 2023 and was fully reconstructed starting in December 2023. 

\subsection{Analysis of $\Lambda$ decay} \label{sec:proofofconcept}
\label{sec:PoC_Lambda}

\begin{figure}
    \centering
    \includegraphics[width=0.7\linewidth]{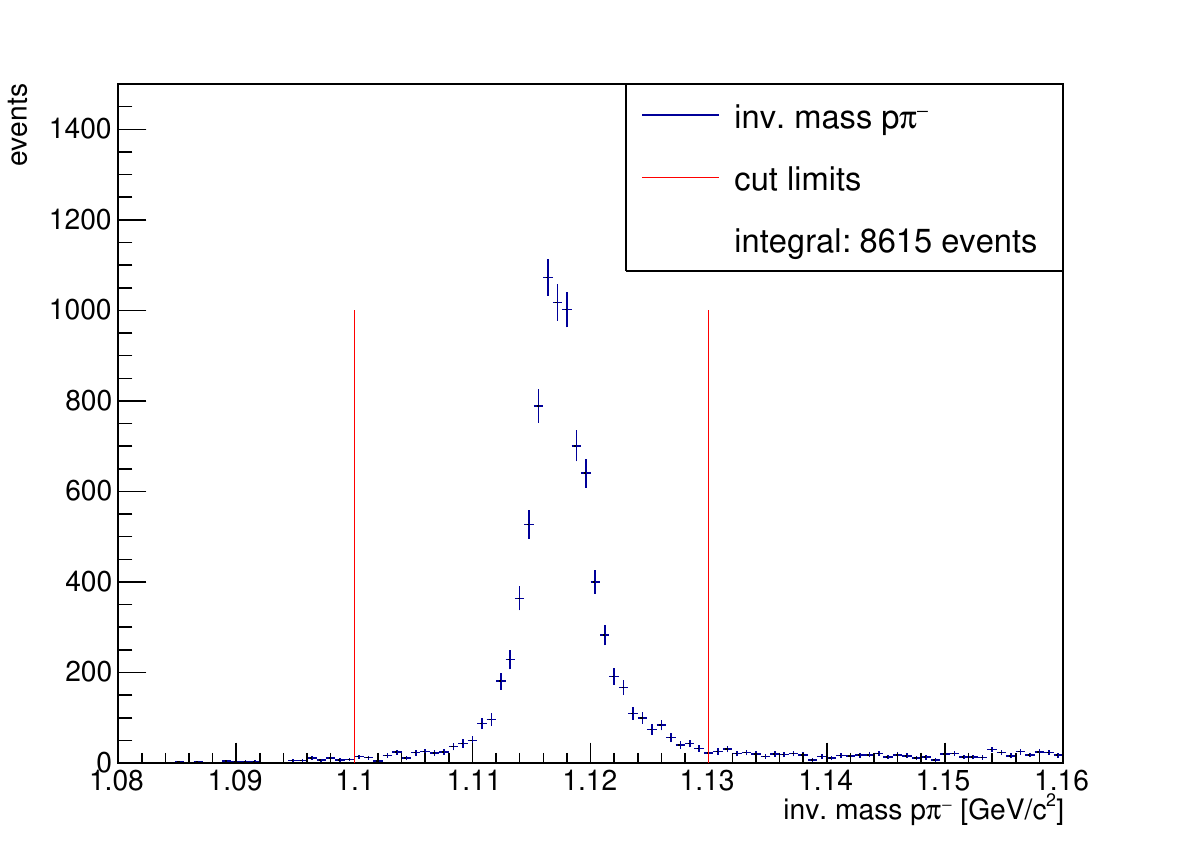}
    \caption{Invariant mass distribution of $p\pi^-$ for $\gamma p \rightarrow K^+ p\pi^-$ events reconstructed from the ``initial subset", representing about 5\% of the GlueX-II 2023 running period.}
    \label{fig:lambdaMass}
\end{figure}

From the initial subset about 8.6k $\Lambda$ events were extracted. 
They were reconstructed in the decay $\Lambda\rightarrow p\pi^-$. The resulting $p\pi^-$ invariant mass distribution is shown in Fig.~\ref{fig:lambdaMass}.
The resulting data is analyzed as outlined in Sec.~\ref{sec:method}. \par
\begin{figure}
    \centering
    \includegraphics[width=\linewidth]{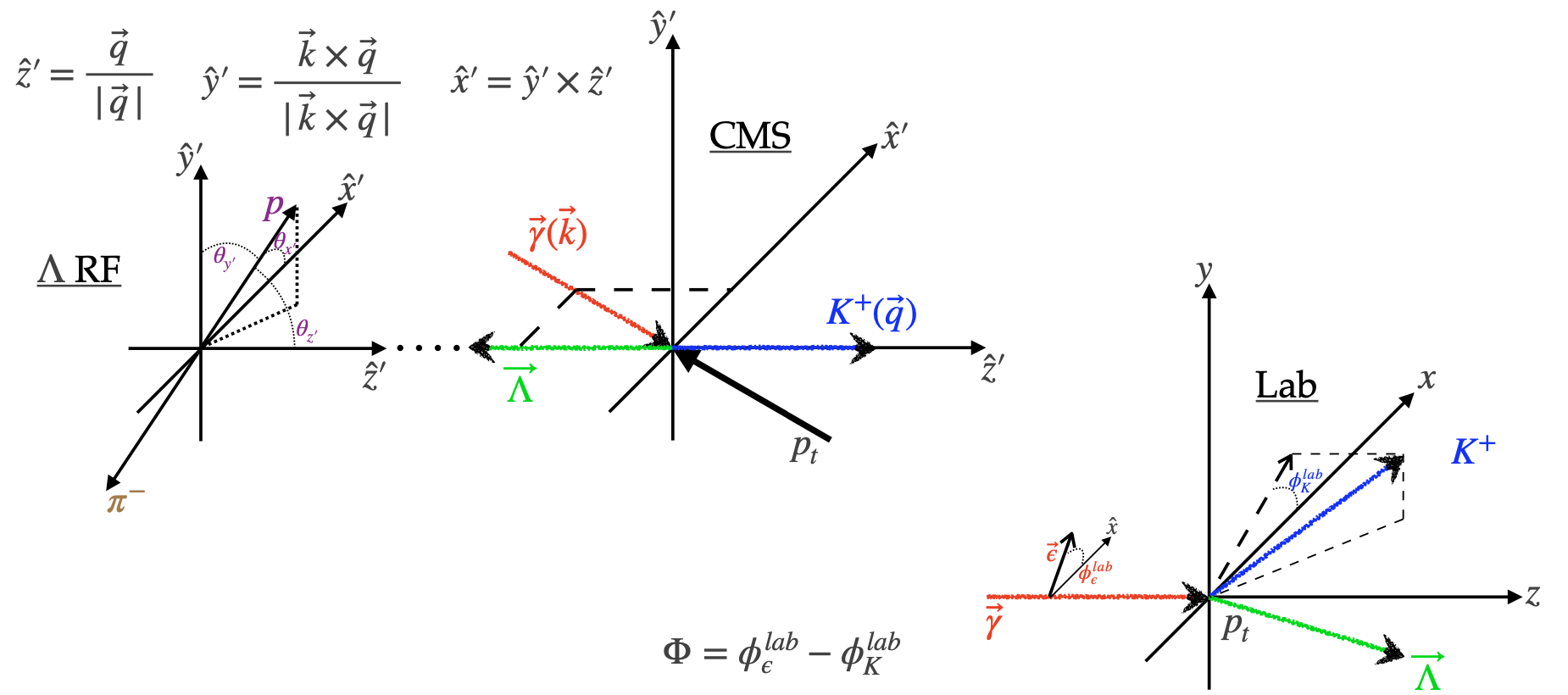}
    \caption{Reference frames used in this analysis}
    \label{fig:frames}
\end{figure}

Figure~\ref{fig:frames} shows the different reference frames used for this analysis.
The fit variables $\theta_{x',y',z'}$ are defined as the projections of the proton direction in the $\Lambda$ rest frame, onto the coordinate system axis. The coordinate system (the so-called primed system) in the $\Lambda$ rest frame is defined by
\begin{align}
    \hat{z}' = \frac{\vec q}{|\vec q|} \hspace{1cm} \hat{y}' = \frac{\vec k \times \vec q}{|\vec k \times \vec q|} \hspace{1cm} \hat{x}' = \hat{y}' \times \hat{z}'
\end{align}
where $\vec q$ denotes the momentum vector of the $K^+$ in the centre-of-mass frame (CMS) and $\vec k$ denotes the momentum vector of the beam photon in the CMS. \par

The extended maximum likelihood can be written as
\begin{align}
    \ln \mathscr{L} = \sum_{i=1}^N\ln I(\Phi,\theta_{x',y',z'}) - \int\text{d}\Omega I(\Phi,\theta_{x',y',z'})\eta(\Phi,\theta_{x',y',z'})
\end{align}
where $I(\Phi,\theta_{x',y',z'})$ is the intensity defined in Eq.~\eqref{eq:intensity} and $\eta(\Phi,\theta_{x',y',z'})$ denotes the detector acceptance which has to be taken into account. The sum is over all $N$ events in the data, while the integral is evaluated by summing over phase-space MC events which have been passed through a Geant4 based simulation of the whole GlueX detector setup and then reconstructed and analyzed like real data. The likelihood function is used in a Markov Chain Monte Carlo (MCMC) parameter optimization process. Instead of trying to minimize this multi-dimensional likelihood it is numerically explored.\par
\begin{figure}
    \centering
    \includegraphics[width=\linewidth]{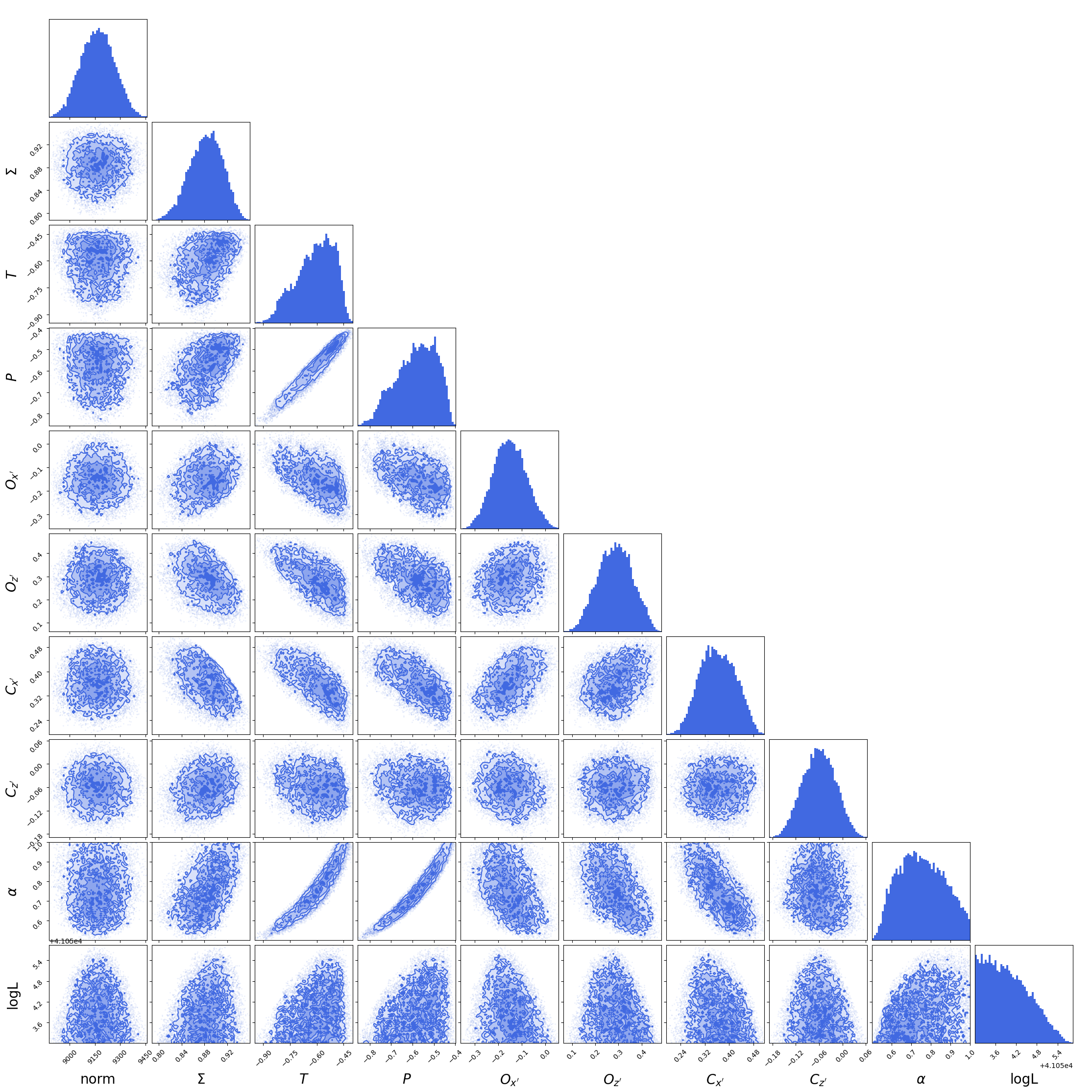}
    \caption{Corner plot visualising the result of the Markov Chain Monte Carlo parameter estimation to determine $\alpha$ together with the seven polarization observables $\Sigma$ to $C_{z'}$. The overall normalization is denoted by {\it norm}.}
    \label{fig:corner}
\end{figure}

The result is shown in Fig.~\ref{fig:corner}.
The mean and standard deviation extracted from the $\alpha$ posterior distribution is
\begin{align*}
    \alpha=0.755 \pm 0.115
\end{align*}
which shows very good agreement with the previous data, albeit with large statistical uncertainty. However, it shows that the proposed method works very well.

\FloatBarrier

\subsection{Analysis of $\rho$ decay for Degree of Circular Polarization}
\label{sec:PoC_rho}

This is essentially an extension of measurements of polarized $\rho$ SDMEs published by GlueX \cite{GlueX:2023fcq} by including circular polarization and constraining the intensity function with the Partial Wave formalism. Indeed conceptually it is similar to the proposed measurement of $\alpha$, whereby the circular polarized intensities are predicted from the linear polarized intensities providing sensitivity to the degree of polarization from the measured decay amplitudes.

Following \cite{Mathieu:2019fts} the intensity is written as,

\begin{align}\label{eq:def_Ipre}
    \mathcal{I}(\Omega,\Phi) = \mathcal{I}^0(\Omega)
    &-P^\gamma_L \mathcal{I}^1(\Omega)\cos(2\Phi)
    -P^\gamma_L \mathcal{I}^2(\Omega) \sin(2\Phi)
    -P^\gamma_C \mathcal{I}^3(\Omega),
\end{align}

The polarized intensity functions can be expanded in moments of Spherical Harmonics which are then related to the contributing partial waves, which should be dominated by $P$ waves for $\rho$ decays. This results in an over-constrained set of relationships allowing us to treat the polarization degrees as unknown parameters when performing fits of Eq.~\eqref{eq:def_Ipre} to the data in terms of the partial waves. See Appendix~\ref{App:rho} for technical details.

\begin{figure}
    \centering
    \includegraphics[width=0.8\linewidth]{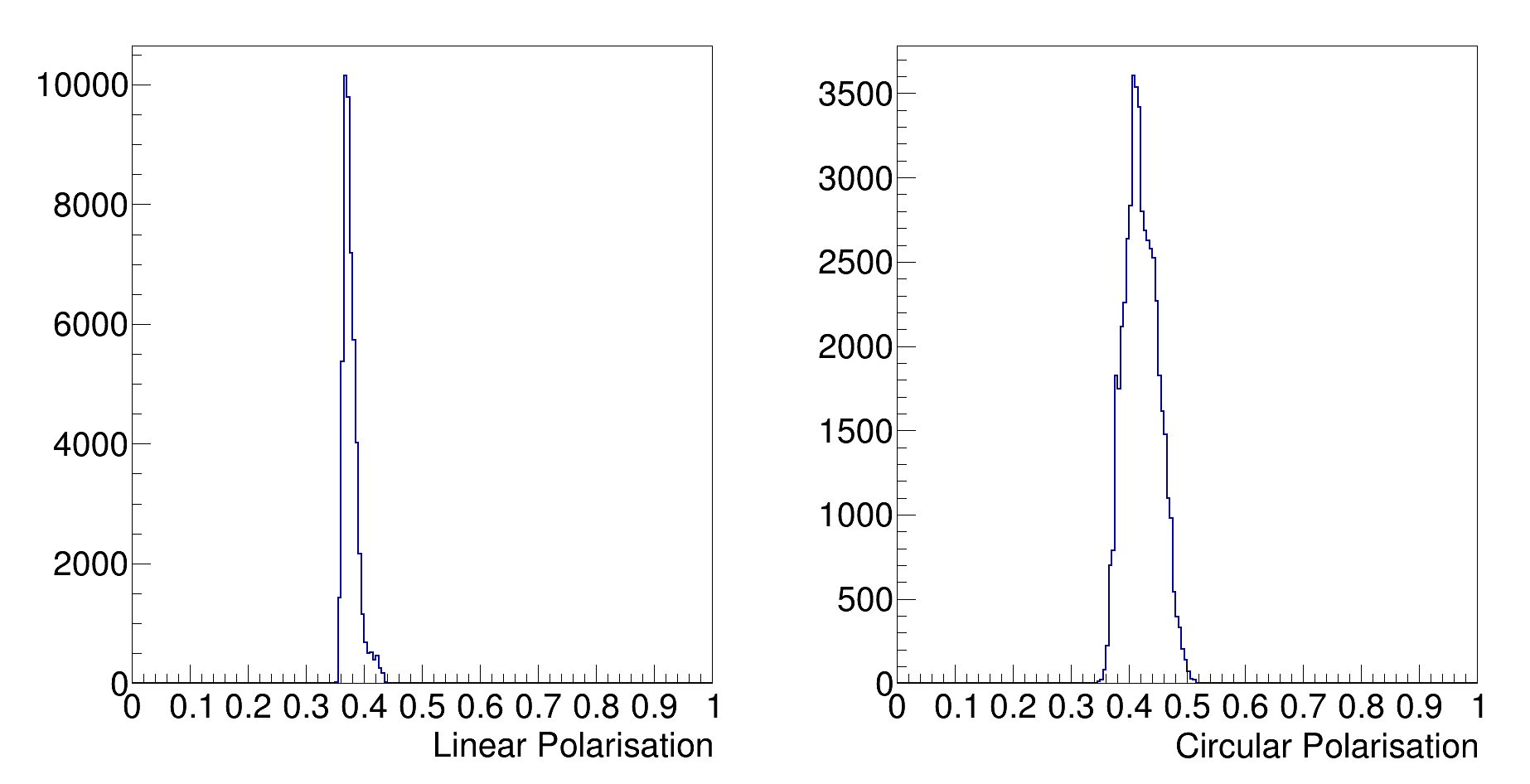}
    \caption{Degrees of linear and circular polarization extracted in preliminary fits to 2023 data. The data distributions show MCMC samples. Numerically we determine $P^\gamma_L = 0.375\pm0.008$~(stat) and $P^\gamma_C=0.423\pm0.027$~(stat), independent of any other polarimetry technique. Systematic uncertainties must yet be studied.}
    \label{fig:resultPLinCirc}
\end{figure}

For the same initial subset of GlueX\nobreakdash-II runs, we again perform MCMC sampling of the log likelihood, but now using the intensity of Eq.~\eqref{eq:def_Ipre} and with the polarization degrees left as free parameters. We select the high t-region ($>$\SI{0.5}{\GeV}), where we expect sensitivity to be largest and fit 390k random subtracted events. The results of this preliminary analysis look very promising. The extracted amplitude values agree with expectations from the $\rho$ SDMEs and are shown in Appendix~\ref{App:rho}, while the resulting polarizations are indeed consistent with their expected values. The results are shown in Fig.~\ref{fig:resultPLinCirc}, where cuts have been placed. Using this method statistical uncertainties on the degrees of polarization will be very small ($<1\%$). Systematic uncertainties will be well understood through measuring the polarizations in many t and invariant mass bins, as well as using other channels, such as $\omega$ production.

It would also be possible to use this reaction to give a simple asymmetry with which to monitor the relative beam polarization on a run-by-run basis, without needing detailed analysis or acceptance corrections. The helicity dependent intensity $\mathcal{I}^3(\Omega)$ contains spherical harmonic terms which flip with the beam helicity. In terms of the $\rho$ SDME elements we can express a 2D asymmetry :

\begin{align*}
    \mathcal{A}_{3} =& \frac{\mathcal{I}(\Omega,h=+1)-\mathcal{I}(\Omega,h=-1)}{\mathcal{I}_{3}(\Omega,h=+1)+\mathcal{I}_{3}(\Omega,h=-1)} = \frac{\mathcal{I}_{3}(\Omega)}{\mathcal{I}_{0}(\Omega)}\\
    \approx&\frac{3}{4\pi}(
    \sqrt{2}\Im{\rho^{3}_{10}}\sin{2\theta}\sin{\phi}
    +\Im{\rho^{3}_{1-1}}\sin^{2}{\theta}\sin{2\phi})/ \mathcal{I}_{0}(\Omega)
\end{align*}

\begin{figure}
    \centering
    \subfloat[]{\includegraphics[width=0.5\linewidth]{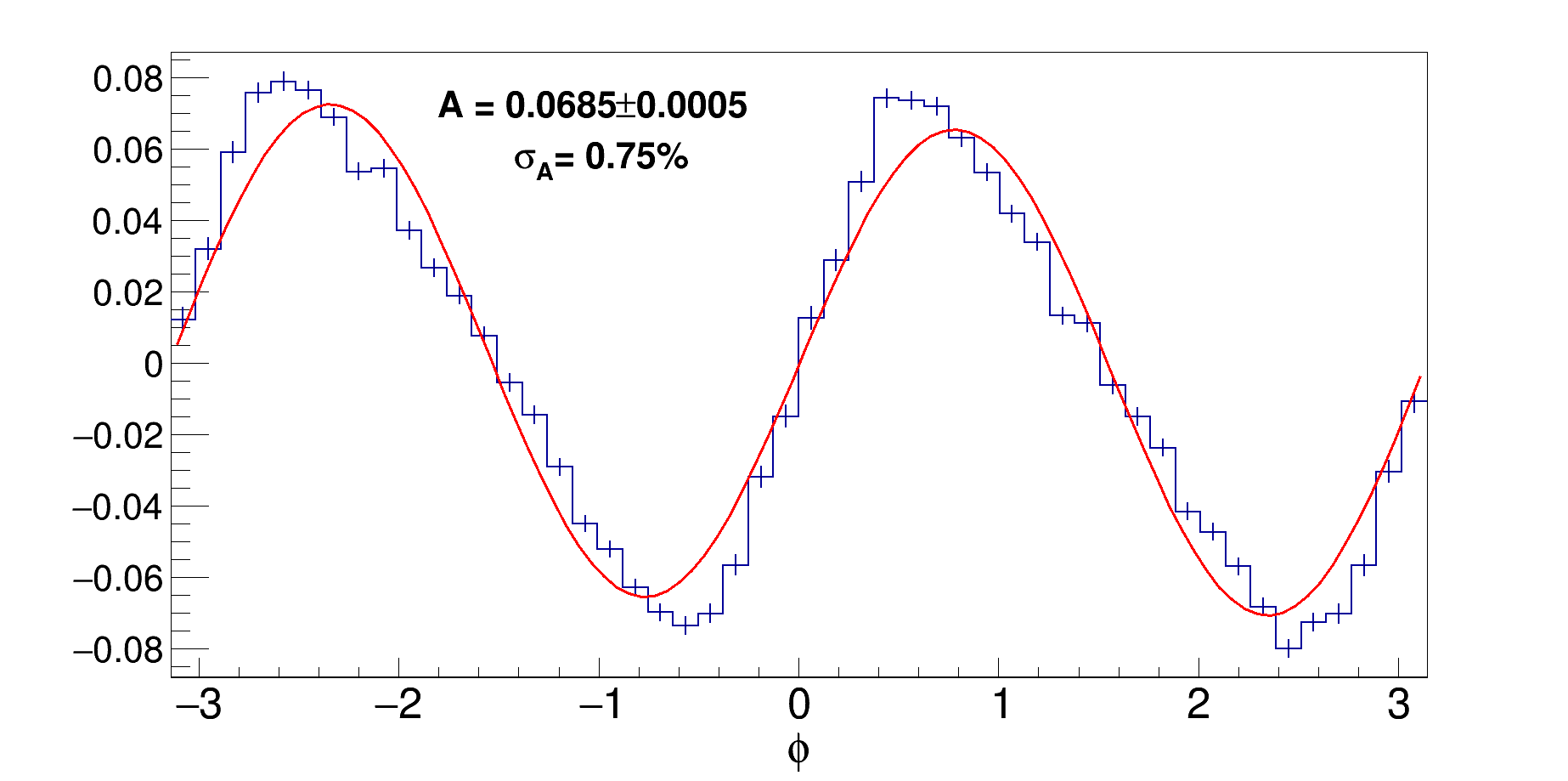}} 
    \subfloat[]{\includegraphics[width=0.5\linewidth]{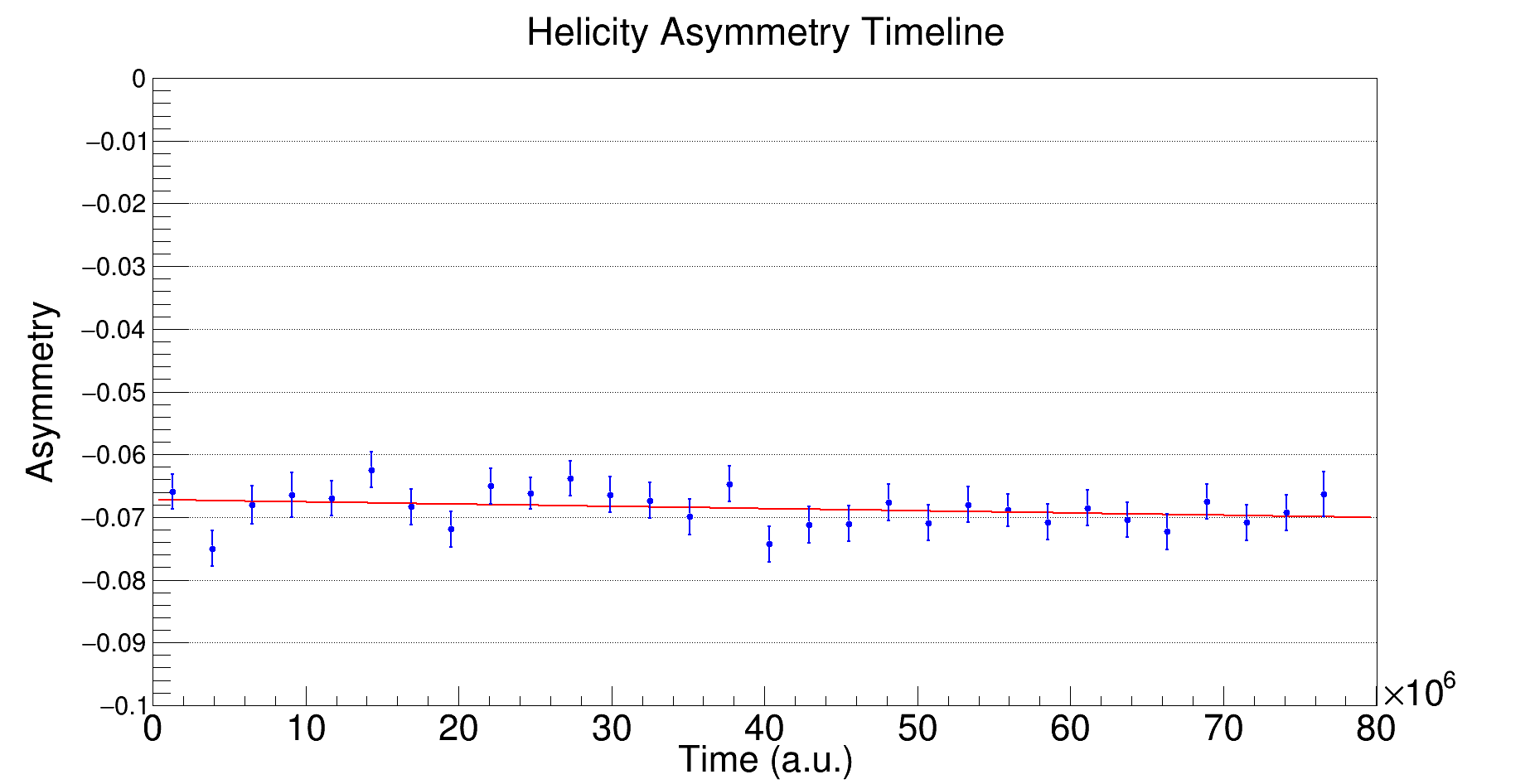}}
    \caption{(a) Helicity asymmetry $\mathcal{A}_{3}(\phi)$ from $\rho$ production using the 30 reconstructed runs from 2023. This asymmetry is proportional to the electron beam polarization. (b) The time dependence of the amplitude of the helicity asymmetry demonstrating sensitivity to changes in the polarization.}
    \label{fig:resultRhoAsymmetry}
\end{figure}

We integrate this over $\cos(\theta)$ for +ve and -ve values separately, projecting onto $\phi$, flipping $\phi$ in the -ve case. The resulting $\mathcal{A}_{3}(\phi)$, for the initial sample of 30 runs, is shown in Fig.~\ref{fig:resultRhoAsymmetry}(a). It can be approximated by a $\sin(2\phi)$ function due to the dominance of the $\rho^{3}_{1-1}$ SDME. The asymmetry amplitude is proportional to the beam polarization. Fitting the integral we get an amplitude of the helicity asymmetry of $A\approx6.85\pm0.05\times10^{-3}$ for beam polarization of $53\%$ (Fig.~\ref{fig:resultRhoAsymmetry}(a)).  This is a precision on the circular polarization of $dP^\gamma_{C}/P^\gamma_{C}=0.73\%$.  Such a precision corresponds to $dP^\gamma_{C}/P^\gamma_{C}=4.0\%$ per 2-hour run.  With a polarization of $75\%$ the uncertainty would decrease by $0.53/0.75$ to become $dP^\gamma_{C}/P^\gamma_{C}=2.7\%$ per run. 
Integrated over the GlueX\nobreakdash-II running, the statistical uncertainty becomes negligible.
Fig.~\ref{fig:resultRhoAsymmetry}(b) shows the amplitude of the helicity asymmetry plotted for each run in the initial sample.   This shows that we have the statistical power to be sensitive to a time dependence of the polarization on typical multi-day timescales which has been previously observed~\cite{Zec:2024iky}.

\begin{figure}
    \centering
    \includegraphics[width=0.8\linewidth]{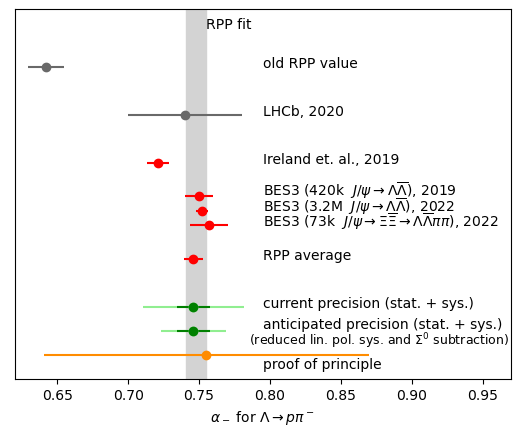}
    \caption{Our proof-of-concept result in comparison to previous data. Also shown is a projection for the achievable statistical precision of the proposal as described in Sec.~\ref{sec:stats} (dark green), as well as our projection for the combined systematic and statistical precision anticipated for this measurement (c.f. Sec.~\ref{sec:totalsys}). The old RPP value and the LHCb result are not used for calculating the RPP average and fit.}
    \label{fig:resultSummary}
\end{figure}

\FloatBarrier

\section{Experiment}
\label{sec:experiment}

We plan to run alongside the remaining allocated beam time for GlueX\nobreakdash-II. We do not require any changes to the setup of the experiment.

\subsection{Statistical Uncertainty}
\label{sec:stats}

Here we use the ``initial subset" of data to project a final statistical uncertainty for the whole of GlueX\nobreakdash-II running.
Recall that this is 30 of 600 runs from 2023 that had an endpoint circular polarization of $\sim$50\% and gave 8.6k clean $\Lambda\rightarrow p\pi^-$ events.
The beam time was split about 100 to 500 into periods of 50\% and 70\% polarization respectively, so we project $\sim$28k events with 50\% polarization, and
$\sim$115k events with 70\% polarization. 
Assuming that the higher polarization goes directly into our statistical precision we have already acquired  an effective 160k events (115k*70\%/50\%) compared to our 30 run subset.
GlueX\nobreakdash-II still has about 220 days of data taking left, which is about 3.3 times more than what we had in the 2023 running. Assuming that we get 80\% electron beam polarization this is effectively about 3.8 times more data that we currently have on tape. \par

In total we project about 28k+160k+700k effective events for GlueX\nobreakdash-II running.  Note, these are not total expected events, but scaled up by polarization to compare to our 30 runs used in the proof-of-principle. That means, in total we expect more than 100 times more compared to our small 8.6k event subset. This would result in 10 times smaller statistical uncertainties compared to what we extracted in the proof-of-principle. This estimate is shown in Fig.~\ref{fig:resultSummary} as projection. \par

The estimate of the statistical accuracy expected for this proposal will be competitive to the current estimate for $\alpha$ as listed in the RPP. Given that we are using a completely different method from BESIII and a much improved methodology over Ireland et al. we will make a crucial contribution to resolve the remaining tension in the data.

\subsection{Systematic Uncertainty}

\subsubsection{Polarimetry}
As described in Sec.~\ref{sec:polarimetry_circ}, a careful analysis of polarization data from other halls and physics data recorded with the GlueX spectrometer will allow us to achieve a polarization systematic uncertainty in Hall D of $\sim2\%$ on the circular polarization.  The linear polarization has already been determined to $\sim2\%$ in previous running periods. \par
To study the effects of these on the extraction of $\alpha_-$, MC simulations were used. The data sample was generated according to the observables extracted in the proof-of-principle and with a nominal circular and linear polarization of 0.65 and 0.38, respectively, and then analyzed with polarizations that were 1$\sigma$ higher. The impact on $\alpha_-$ was determined to be negligible for the circular polarization and adding about $4\%$ systematic uncertainty in the case of the linear polarization. \par
As shown in Sec.~\ref{sec:PoC_rho}, the analysis of specific physics reactions can be used to determine the polarization. We believe that we can use this to improve the systematic uncertainty on the linear polarization. Furthermore, the effect of the linear polarization is only so large because of the values and correlations for polarization observables $T$ and $P$. In our proof-of-principle we do not have enough data to bin in momentum transfer $-t$, but it is expected that the polarization observables will change depending on $-t$. 
By binning in $-t$ we would be able to study the extraction with different sensitivities to the linear polarization allowing us to develop strategies to minimize this uncertainty.
If necessary we could choose a range in $-t$ for our final analysis that allows us to reduce the systematic uncertainty on the linear polarization in exchange for some statistical precision. \par
As such, the numbers quoted above are a conservative upper limit for the systematic uncertainty of the polarization on our final result and we are cautiously optimistic that we will be able to halve them in our final result.

\subsubsection{Acceptance}
The measurement of $\alpha_-$ relies on an acceptance correction which is performed by using MC simulations. The GlueX collaboration developed an excellent understanding of its detector and accurate MC simulations are achieved through the Geant4 based detector simulation package {\it hdgeant4}. Previous publications indicate that we can assume a systematic effect from the acceptance correction of less than 2\%.

\subsubsection{Background contamination}
One potential large background for $\Lambda$ photoproduction comes from the $\Sigma^0$ photoproduction. The $\Sigma^0\rightarrow\Lambda\gamma$ decay produces soft photons which can be missed due to detector thresholds. Ref.~\cite{Boyarski:1969iy} indicates slightly larger $\Lambda$ than $\Sigma^0$ production cross-sections at $E_\gamma=\SI{8}{GeV}$. To study the contamination, simulations for $\gamma p\rightarrow K^+\Lambda$ and $\gamma p\rightarrow K^+\Sigma^0$ were produced and analyzed. We could show that strictly requiring no neutral showers in our detector reduces the $\Sigma^0$ background substantially while preserving most of the signal. Furthermore, one can use the $\phi$-angles of the $K^+$ and $\Lambda$ to improve purity further. A simple requirement that $K^+$ and $\Lambda$ $\phi$-angles are back-to-back reduces the $\Sigma^0$ contamination further, in total to less than 2.5\%. One could also use them as a discriminatory variable to perform a background subtraction (e.g. using sWeights) and improve purity even further. This might actually be the best way to remove the background contamination.\par 
To study the effect of the background on the extraction of $\alpha_-$ we assume a worst case scenario, where we only use the cut-based approach to reduce the $\Sigma^0$ contamination. Assuming that it is produced exactly the same way as the signal from $\Lambda$, it would carry some polarization itself which is passed on to the $\Lambda$ with a factor of $-\frac{1}{3}$ \cite{CLAS:2006pde}, which would then reduce the measured $\Lambda$ polarization and hence dilute our measurement for $\alpha_-$ by about 3\%. We expect to correct for the bias with better than 10\% precision so we conservatively assume 0.3\% systematic uncertainty.\par
In summary, assuming the same amount of $\Lambda/\Sigma^0$ being produced, based on our studies, less than 2.5\% contamination from $\Sigma^0$ production is expected, which we can correct for. Ultimately, we plan to subtract the remaining background which will result in a negligible effect from $\Sigma^0$ contamination. In a worst case scenario where we cannot do the subtraction, it would introduce a systematic bias of about 0.3\% on our extraction of $\alpha_-$.

\subsubsection{Total}\label{sec:totalsys}
A summary of the main systematic uncertainties anticipated for this measurement is provided in Table~\ref{tab:systematics}. For now, we assume that the linear polarization uncertainty as well as $\Sigma^0$ contamination can be taken as upper limits for how much they will ultimately contribute to the $\alpha_-$ measurement. Adding all systematic uncertainties in quadrature, in total we expect a systematic uncertainty of less than $5.4\%$. With the improvements planned for the linear polarization measurement, as well as the $\Sigma^0$ background subtraction, we are optimistic to reach a total systematic uncertainty of $2.8\%$. Both numbers are included in Figure~\ref{fig:resultSummary} to show the current upper limit and the anticipated total uncertainty expected for our measurement.
\begin{table}[ht]
    \centering
    \caption{Summary of the systematic uncertainties expected in this measurement.  Column 2 gives the percentage systematic uncertainty on each quantity and column 3 gives the effect on the final result. Column 4 lists the absolute contribution for our measurement based on a value of $\alpha_-=0.75$. The total is calculated for the upper limit as well as our expected systematic uncertainty.}
    \begin{tabular}{l|c|c|c}
          &  \% uncertainty   &  \% contribution & absolute contribution\\ 
        \hline
        Photon beam circular polarization  &  2\% & $<$0.2\% & $<$0.002 \\
        Photon beam linear polarization  &  $<$2\% & $<$4\% & $<$0.03 \\
        Acceptance &  2\% & 2\% & 0.015 \\
        $\Sigma$ contamination & $<$2.5\% & $<$0.3\% & $<$0.002 \\\hline
        Total (current upper limit)&   &  $<$4.5\% & $<$0.034 \\
        Total (anticipated)&   &  2.8\% & 0.021 \\
    \end{tabular}
    \label{tab:systematics}
\end{table}

\section{Additional physics possible with circularly polarized photons in Hall D} \label{sec:additionalphysics}

While the main focus is the weak decay parameter $\alpha_-$ for $\Lambda\rightarrow\pi^-p$, the requests described in this proposal, namely providing circular polarization to Hall D and the mechanisms to measure it, will allow access to other publishable physics. We outline several of these below.

\subsection{$\Sigma^+$ Weak Decay Parameter}
The same methodology as outlined for the measurement of $\alpha_-$ can be used to analyze the weak decay parameter for $\Sigma^+\rightarrow\pi^0p$ in the reaction $\gamma p\rightarrow\Sigma^+K^0_s$. The current RPP value is $-0.982\pm0.014$ and based on a measurement by BESIII ($J/\psi\rightarrow\Sigma^+\bar{\Sigma}^-$) and measurements from the 1970s using recoil polarization measurements. Although all previous measurements agree we would provide a new and methodologically different measurement.

\subsection{Polarized Spin Density Matrix Elements}
As described in Sec.~\ref{sec:polarimetry_circ}, the vector mesons have spin-density matrix elements (SDMEs) which describe how the photon polarization is transferred to the meson and give information on the production mechanisms~\cite{Schilling:1969um}.
The presence of a known amount of circular polarization in the beam will allow the extraction of additional SDMEs in the photon energy range $6 < E_{\gamma} < 12$\,GeV for $\gamma p\to\rho p$, $\gamma p\to\omega p$,  $\gamma p\to\phi p$ and $\gamma p\to J/\psi p$. In the latter case these may provide information on the presence of hadron resonance contributions to the production process.

Similarly, additional SDMEs will be measurable for t-channel baryon production such as $\gamma p\to\pi \Delta$ or $\gamma p\to K^{+} \Lambda (1520)$. 

In this proposal the set of seven polarization observables measured simultaneously with the $\alpha_-$ parameter are essentially an example of such SDME measurements, though written in a different formalism. They will provide an additional publication from the same analysis.

These spin observables allow investigation of the mechanisms which contribute to photoproduction. For example, which particles are exchanged, to what degree is helicity conserved. This then allows and encourages development of more sophisticated and better constrained reaction models.

\subsection{Amplitude Analysis for Meson Spectroscopy}

As circular polarization project out additional SDMEs in meson production reactions, these SDMEs then produce further constraints on contributing partial waves. These additional constraints can act to reduce possible ambiguities, in particular removing a complex conjugate ambiguity, and reduce statistical uncertainties compared to beamtimes without this additional polarization. This would therefore provide extra assistance for the core GlueX program.

\subsection{Timelike Compton Scattering}

The helicity asymmetry in Timelike Compton scattering (TCS), measured in the process $\gamma p\to e^+e^- p$, is particularly useful as it is an observable which is zero when there is no TCS contribution. This angular asymmetry of the decay
leptons accesses the real part of the Compton form factors, important for the Generalized Parton Distribution framework. It has recently been measured for the first time by the CLAS12 collaboration \cite{PhysRevLett.127.262501} and independent measurements at GlueX would provide highly competitive results.

\section{Request}

We request that GlueX be given a share in the beam polarization.
In concrete terms, we request that Hall D be included in the usual calculations to determine the optimal running conditions that maximize polarization in multiple halls simultaneously.
The optimization is done changing the the linac energies individually and the Wien angle~\cite{Higinbotham:2009ze} and is typically able to find a configuration which provides polarization to all halls.

\section{Summary}
In this document we outline how using the existing GlueX apparatus and approved beam time we can measure weak decay constant $\alpha_-$ for the decay $\Lambda\rightarrow p\pi^-$.
This measurement will be able to resolve a significant discrepancy in the literature.
We ask for running in parallel with GlueX and to be included in the usual calculations to determine the optimal running conditions that maximize polarization in multiple halls simultaneously.


\printbibliography 

@article{BESIII:2022qax,
    author = "Ablikim, M. and others",
    collaboration = "BESIII",
    title = "{Precise Measurements of Decay Parameters and $CP$ Asymmetry with Entangled $\Lambda-\bar{\Lambda}$ Pairs Pairs}",
    eprint = "2204.11058",
    archivePrefix = "arXiv",
    primaryClass = "hep-ex",
    doi = "10.1103/PhysRevLett.129.131801",
    journal = "Phys. Rev. Lett.",
    volume = "129",
    number = "13",
    pages = "131801",
    year = "2022"
}

@article{A2:2024ydg,
    author = "Afzal, F. and others",
    collaboration = "A2",
    title = "{First Measurement Using Elliptically Polarized Photons of the Double-Polarization Observable E for \ensuremath{\gamma}p\textrightarrow{}p\ensuremath{\pi}0 and \ensuremath{\gamma}p\textrightarrow{}n\ensuremath{\pi}+}",
    eprint = "2402.05531",
    archivePrefix = "arXiv",
    primaryClass = "nucl-ex",
    doi = "10.1103/PhysRevLett.132.121902",
    journal = "Phys. Rev. Lett.",
    volume = "132",
    number = "12",
    pages = "121902",
    year = "2024"
}

@article{olsen,
  title = {Photon and Electron Polarization in High-Energy Bremsstrahlung and Pair Production with Screening},
  author = {Olsen, H. and Maximon, L. C.},
  journal = {Phys. Rev.},
  volume = {114},
  issue = {3},
  pages = {887-904},
  numpages = {0},
  year = {1959},
  publisher = {American Physical Society},
  doi = {10.1103/PhysRev.114.887},
  url = {https://link.aps.org/doi/10.1103/PhysRev.114.887}
}

@article{GlueX:2023fcq,
    author = "Adhikari, S. and others",
    collaboration = "GlueX",
    title = "{Measurement of spin-density matrix elements in \ensuremath{\rho}(770) production with a linearly polarized photon beam at E\ensuremath{\gamma}=8.2\textendash{}8.8~GeV}",
    eprint = "2305.09047",
    archivePrefix = "arXiv",
    primaryClass = "nucl-ex",
    reportNumber = "JLAB-PHY-23-3838",
    doi = "10.1103/PhysRevC.108.055204",
    journal = "Phys. Rev. C",
    volume = "108",
    number = "5",
    pages = "055204",
    year = "2023"
}

@article{GlueX:2017zoo,
    author = "Al Ghoul, H. and others",
    collaboration = "GlueX",
    title = "{Measurement of the beam asymmetry $\Sigma$ for $\pi^0$ and $\eta$ photoproduction on the proton at $E_\gamma = 9$ GeV}",
    eprint = "1701.08123",
    archivePrefix = "arXiv",
    primaryClass = "nucl-ex",
    reportNumber = "JLAB-PHY-17-2403",
    doi = "10.1103/PhysRevC.95.042201",
    journal = "Phys. Rev. C",
    volume = "95",
    number = "4",
    pages = "042201",
    year = "2017"
}

@article{Barker:1975bp,
    author = "Barker, I. S. and Donnachie, A. and Storrow, J. K.",
    title = "{Complete Experiments in Pseudoscalar Photoproduction}",
    reportNumber = "DL/P 232",
    doi = "10.1016/0550-3213(75)90049-8",
    journal = "Nucl. Phys. B",
    volume = "95",
    pages = "347--356",
    year = "1975"
}

@article{Dugger:2017zoq,
    author = "Dugger, M. and others",
    title = "{Design and construction of a high-energy photon polarimeter}",
    eprint = "1703.07875",
    archivePrefix = "arXiv",
    primaryClass = "physics.ins-det",
    doi = "10.1016/j.nima.2017.05.026",
    journal = "Nucl. Instrum. Meth. A",
    volume = "867",
    pages = "115--127",
    year = "2017"
}

@article{Grames:2004mk,
    author = "Grames, J. M. and others",
    title = "{Unique electron polarimeter analyzing power comparison and precision spin-based energy measurement}",
    reportNumber = "JLAB-ACO-04-11",
    doi = "10.1103/PhysRevSTAB.7.042802",
    journal = "Phys. Rev. ST Accel. Beams",
    volume = "7",
    pages = "042802",
    year = "2004",
    note = "[Erratum: Phys.Rev.ST Accel.Beams 13, 069901 (2010)]"
}

@article{Higinbotham:2009ze,
    author = "Higinbotham, D. W.",
    editor = "Crabb, Donald G. and Day, Donal B. and Liuti, Simonetta and Zheng, Xiaochao and Poelker, Matt and Prok, Yelena",
    title = "{Electron Spin Precession at CEBAF}",
    eprint = "0901.4484",
    archivePrefix = "arXiv",
    primaryClass = "physics.acc-ph",
    reportNumber = "JLAB-PHY-09-942",
    doi = "10.1063/1.3215753",
    journal = "AIP Conf. Proc.",
    volume = "1149",
    number = "1",
    pages = "751--754",
    year = "2009"
}

@article{Ireland:2019uja,
    author = {Ireland, D. G. and D\"oring, M. and Glazier, D. I. and Haidenbauer, J. and Mai, M. and Murray-Smith, R. and R\"onchen, D.},
    title = "{Kaon Photoproduction and the $\Lambda$ Decay Parameter $\alpha_-$}",
    eprint = "1904.07616",
    archivePrefix = "arXiv",
    primaryClass = "nucl-ex",
    doi = "10.1103/PhysRevLett.123.182301",
    journal = "Phys. Rev. Lett.",
    volume = "123",
    number = "18",
    pages = "182301",
    year = "2019"
}

@article{Mathieu:2019fts,
    author = "Mathieu, V. and Albaladejo, M. and Fern\'andez-Ram\'\i{}rez, C. and Jackura, A. W. and Mikhasenko, M. and Pilloni, A. and Szczepaniak, A. P.",
    collaboration = "JPAC",
    title = "{Moments of angular distribution and beam asymmetries in $\eta\pi^0$ photoproduction at GlueX}",
    eprint = "1906.04841",
    archivePrefix = "arXiv",
    primaryClass = "hep-ph",
    reportNumber = "JLAB-THY-19-2958",
    doi = "10.1103/PhysRevD.100.054017",
    journal = "Phys. Rev. D",
    volume = "100",
    number = "5",
    pages = "054017",
    year = "2019"
}

@article{Schilling:1969um,
    author = "Schilling, K. and Seyboth, P. and Wolf, G. E.",
    title = "{On the Analysis of Vector Meson Production by Polarized Photons}",
    reportNumber = "SLAC-PUB-0683",
    doi = "10.1016/0550-3213(70)90070-2",
    journal = "Nucl. Phys. B",
    volume = "15",
    pages = "397--412",
    year = "1970",
    note = "[Erratum: Nucl.Phys.B 18, 332 (1970)]"
}

@article{Sandorfi:2010uv,
    author = "Sandorfi, A. M. and Hoblit, S. and Kamano, H. and Lee, T. -S. H.",
    title = "{Determining pseudoscalar meson photo-production amplitudes from complete experiments}",
    eprint = "1010.4555",
    archivePrefix = "arXiv",
    primaryClass = "nucl-th",
    reportNumber = "JLAB-THY-10-1268",
    doi = "10.1088/0954-3899/38/5/053001",
    journal = "J. Phys. G",
    volume = "38",
    pages = "053001",
    year = "2011"
}

@article{Boyarski:1969iy,
    author = "Boyarski, A. and Bulos, F. and Busza, W. and Diebold, Robert E. and Ecklund, Stanley D. and Fischer, G. E. and Murata, Y. and Rees, John R. and Richter, Burton and Williams, W. S. C.",
    title = "{PHOTOPRODUCTION OF K+ LAMBDA AND K+ SIGMA0 FROM HYDROGEN FROM 5-GeV to 16-Gev}",
    reportNumber = "SLAC-PUB-0552",
    doi = "10.1103/PhysRevLett.22.1131",
    journal = "Phys. Rev. Lett.",
    volume = "22",
    pages = "1131--1133",
    year = "1969"
}

@article{Sakharov:1967dj,
    author = "Sakharov, A. D.",
    title = "{Violation of CP Invariance, C asymmetry, and baryon asymmetry of the universe}",
    doi = "10.1070/PU1991v034n05ABEH002497",
    journal = "Pisma Zh. Eksp. Teor. Fiz.",
    volume = "5",
    pages = "32--35",
    year = "1967"
}

@article{LHCb:2016yco,
    author = "Aaij, R. and others",
    collaboration = "LHCb",
    title = "{Measurement of matter-antimatter differences in beauty baryon decays}",
    eprint = "1609.05216",
    archivePrefix = "arXiv",
    primaryClass = "hep-ex",
    reportNumber = "CERN-EP-2016-212, LHCB-PAPER-2016-030",
    doi = "10.1038/nphys4021",
    journal = "Nature Phys.",
    volume = "13",
    pages = "391--396",
    year = "2017"
}

@article{BESIII:2018cnd,
    author = "Ablikim, M. and others",
    collaboration = "BESIII",
    title = "{Polarization and Entanglement in Baryon-Antibaryon Pair Production in Electron-Positron Annihilation}",
    eprint = "1808.08917",
    archivePrefix = "arXiv",
    primaryClass = "hep-ex",
    doi = "10.1038/s41567-019-0494-8",
    journal = "Nature Phys.",
    volume = "15",
    pages = "631--634",
    year = "2019"
}

@article{Cronin:1963zb,
    author = "Cronin, J. W. and Overseth, O. E.",
    title = "{Measurement of the decay parameters of the Lambda0 particle}",
    doi = "10.1103/PhysRev.129.1795",
    journal = "Phys. Rev.",
    volume = "129",
    pages = "1795--1807",
    year = "1963"
}

@article{Overseth:1967zz,
    author = "Overseth, O. E. and Roth, R. F.",
    title = "{Time Reversal Invariance in Lambda0 Decay}",
    doi = "10.1103/PhysRevLett.19.391",
    journal = "Phys. Rev. Lett.",
    volume = "19",
    pages = "391--393",
    year = "1967"
}

@article{Dauber:1969hg,
    author = "Dauber, P. M. and Berge, J. P. and Hubbard, J. R. and Merrill, D. W. and Muller, R. A.",
    title = "{Production and decay of cascade hyperons}",
    doi = "10.1103/PhysRev.179.1262",
    journal = "Phys. Rev.",
    volume = "179",
    pages = "1262--1285",
    year = "1969"
}

@article{Astbury:1975hn,
    author = "Astbury, P. and others",
    title = "{Measurement of the Differential Cross-Section and the Spin-Correlation Parameters P,A, and R in the Backward Peak of pi- p --\ensuremath{>} K0 Lambda at 5-GeV/c}",
    reportNumber = "Print-75-0675 (CERN)",
    doi = "10.1016/0550-3213(75)90054-1",
    journal = "Nucl. Phys. B",
    volume = "99",
    pages = "30--52",
    year = "1975"
}

@article{Cleland:1972fa,
    author = "Cleland, W. E. and Conforto, G. and Eaton, G. H. and Gerber, H. J. and Reinharz, M. and Gautschi, A. and Heer, E. and Revillard, C. and Von Dardel, G.",
    title = "{A measurement of the beta-parameter in the charged nonleptonic decay of the lambda0 hyperon}",
    doi = "10.1016/0550-3213(72)90544-5",
    journal = "Nucl. Phys. B",
    volume = "40",
    pages = "221--254",
    year = "1972"
}

@article{ParticleDataGroup:2018ovx,
    author = "Tanabashi, M. and others",
    collaboration = "Particle Data Group",
    title = "{Review of Particle Physics}",
    doi = "10.1103/PhysRevD.98.030001",
    journal = "Phys. Rev. D",
    volume = "98",
    number = "3",
    pages = "030001",
    year = "2018"
}

@article{LHCb:2020iux,
    author = "Aaij, R. and others",
    collaboration = "LHCb",
    title = "{Measurement of the $\Lambda^0_b\rightarrow J/\psi\Lambda$ angular distribution and the $\Lambda^0_b$ polarisation in $pp$ collisions}",
    eprint = "2004.10563",
    archivePrefix = "arXiv",
    primaryClass = "hep-ex",
    reportNumber = "LHCb-PAPER-2020-005, CERN-EP-2020-051",
    doi = "10.1007/JHEP06(2020)110",
    journal = "JHEP",
    volume = "06",
    pages = "110",
    year = "2020"
}

@article{BESIII:2021ypr,
    author = "Ablikim, M. and others",
    collaboration = "BESIII",
    title = "{Probing CP symmetry and weak phases with entangled double-strange baryons}",
    eprint = "2105.11155",
    archivePrefix = "arXiv",
    primaryClass = "hep-ex",
    doi = "10.1038/s41586-022-04624-1",
    journal = "Nature",
    volume = "606",
    number = "7912",
    pages = "64--69",
    year = "2022"
}

@article{PhysRevLett.127.262501,
  title = {First Measurement of Timelike Compton Scattering},
  author = {Chatagnon, P. and Niccolai, S. and Stepanyan, S. and Amaryan, M. J. and Angelini, G. and Armstrong, W. R. and Atac, H. and Ayerbe Gayoso, C. and Baltzell, N. A. and Barion, L. and Bashkanov, M. and Battaglieri, M. and Bedlinskiy, I. and Benmokhtar, F. and Bianconi, A. and Biondo, L. and Biselli, A. S. and Bondi, M. and Boss\`u, F. and Boiarinov, S. and Briscoe, W. J. and Brooks, W. K. and Bulumulla, D. and Burkert, V. D. and Carman, D. S. and Carvajal, J. C. and Caudron, M. and Celentano, A. and Chetry, T. and Ciullo, G. and Clark, L. and Cole, P. L. and Contalbrigo, M. and Costantini, G. and Crede, V. and D'Angelo, A. and Dashyan, N. and Defurne, M. and De Vita, R. and Deur, A. and Diehl, S. and Djalali, C. and Dupr\'e, R. and Egiyan, H. and Ehrhart, M. and El Alaoui, A. and El Fassi, L. and Elouadrhiri, L. and Fegan, S. and Fersch, R. and Filippi, A. and Gavalian, G. and Ghandilyan, Y. and Gilfoyle, G. P. and Girod, F. X. and Glazier, D. I. and Golubenko, A. A. and Gothe, R. W. and Gotra, Y. and Griffioen, K. A. and Guidal, M. and Guo, L. and Hakobyan, H. and Hattawy, M. and Hayward, T. B. and Heddle, D. and Hobart, A. and Holtrop, M. and Hyde, C. E. and Ilieva, Y. and Ireland, D. G. and Isupov, E. L. and Jo, H. S. and Joo, K. and Kabir, M. L. and Keller, D. and Khachatryan, G. and Khanal, A. and Kim, A. and Kim, W. and Kripko, A. and Kubarovsky, V. and Kuhn, S. E. and Lanza, L. and Leali, M. and Lee, S. and Lenisa, P. and Livingston, K. and MacGregor, I. J. D. and Marchand, D. and Marsicano, L. and Mascagna, V. and McKinnon, B. and McLauchlin, C. and Migliorati, S. and Mirazita, M. and Mokeev, V. and Montgomery, R. A. and Munoz Camacho, C. and Nadel-Turonski, P. and Naidoo, P. and Neupane, K. and O'Connell, T. R. and Osipenko, M. and Ouillon, M. and Pandey, P. and Paolone, M. and Pappalardo, L. L. and Paremuzyan, R. and Pasyuk, E. and Phelps, W. and Pogorelko, O. and Poudel, J. and Price, J. W. and Prok, Y. and Raue, B. A. and Reed, T. and Ripani, M. and Rizzo, A. and Rossi, P. and Rowley, J. and Sabati\'e, F. and Schmidt, A. and Segarra, E. P. and Sharabian, Y. G. and Shirokov, E. V. and Shrestha, U. and Sokhan, D. and Soto, O. and Sparveris, N. and Strakovsky, I. I. and Strauch, S. and Tyler, N. and Tyson, R. and Ungaro, M. and Vallarino, S. and Venturelli, L. and Voskanyan, H. and Vossen, A. and Voutier, E. and Watts, D. P. and Wei, K. and Wei, X. and Wishart, R. and Yale, B. and Zachariou, N. and Zhang, J. and Zhao, Z. W.},
  collaboration = {CLAS Collaboration},
  journal = {Phys. Rev. Lett.},
  volume = {127},
  issue = {26},
  pages = {262501},
  numpages = {7},
  year = {2021},
  month = {Dec},
  publisher = {American Physical Society},
  doi = {10.1103/PhysRevLett.127.262501},
  url = {https://link.aps.org/doi/10.1103/PhysRevLett.127.262501}
}

@article{Zec:2024iky,
    author = "Zec, A. and others",
    title = "{Ultrahigh-precision Compton polarimetry at 2 GeV}",
    eprint = "2402.16135",
    archivePrefix = "arXiv",
    primaryClass = "physics.ins-det",
    reportNumber = "JLAB-PHY-23-3924",
    doi = "10.1103/PhysRevC.109.024323",
    journal = "Phys. Rev. C",
    volume = "109",
    number = "2",
    pages = "024323",
    year = "2024"
}

@article{CLAS:2006pde,
    author = "Bradford, R. K. and others",
    collaboration = "CLAS",
    title = "{First measurement of beam-recoil observables C(x) and C(z) in hyperon photoproduction}",
    eprint = "nucl-ex/0611034",
    archivePrefix = "arXiv",
    reportNumber = "JLAB-PHY-06-587",
    doi = "10.1103/PhysRevC.75.035205",
    journal = "Phys. Rev. C",
    volume = "75",
    pages = "035205",
    year = "2007"
}

\newpage

\appendix
\section{Circular polarization degree when using a diamond radiator} 
\label{app:circpol}
\begin{figure}[H]
    \centering
    \includegraphics[width=0.8\textwidth]{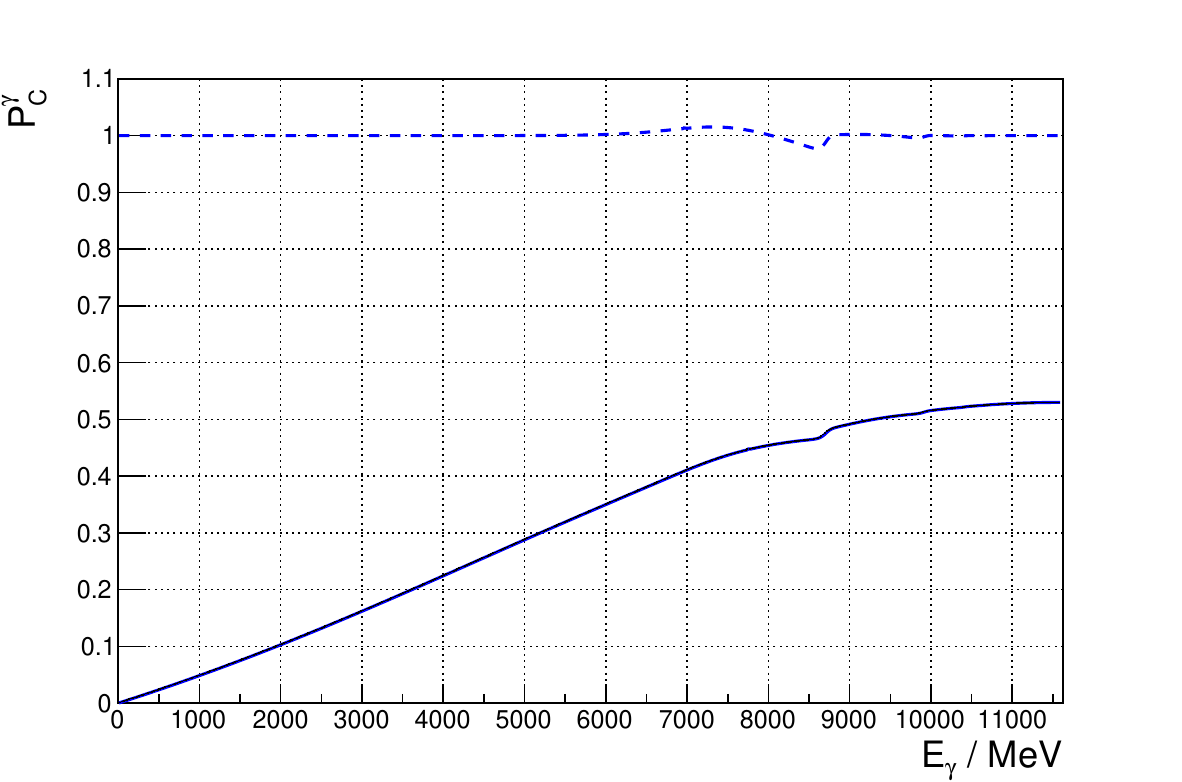}
    \caption{The circular polarization degree as a function of the incoming photon energy when using a diamond radiator (solid line) with the relevant beam, collimator and diamond parameters for the GlueX experiment. The relative deviation between the solid curve and Eq.~\eqref{eq:circpol} are shown with the dashed line at the top.}
    \label{fig:circpol_dia}
\end{figure}

When using a diamond radiator,  the circular polarization degree exhibits a dependence on the diamond lattice, which leads to a decrease of polarization at the position of the coherent peaks \cite{A2:2024ydg}. The size of the decrease depends on the beam, diamond and collimator parameters, as well as the chosen coherent peak position. Fig.~\ref{fig:circpol_dia} shows the calculation for the coherent edge position at 8.6~GeV and an electron polarization degree of $p_e=53$\%  as used in the 2023 running (see Tab.~\ref{tab:polarization}). We find that for our case the effects from the diamond lattice and beam/collimator to have a smaller magnitude than our uncertainty on the circular polarization with an approximate $1.2$\% relative deviation on average in the range of the coherent edge ($8.0$~GeV$<E_\gamma<8.6$~GeV). Therefore we can reliably use Eq.~\eqref{eq:circpol} as a good approximation.

\section{Relationship of S and P waves to the intensity in polarized $\rho$ photoproduction}\label{App:rho}

The intensity for polarized $\rho$ photoproduction can be written as \cite{Mathieu:2019fts},

\begin{align}
    \mathcal{I}(\Omega,\Phi) = \mathcal{I}^0(\Omega)
    &-P^\gamma_L \mathcal{I}^1(\Omega)\cos(2\Phi)
    -P^\gamma_L \mathcal{I}^2(\Omega) \sin(2\Phi)
    -P^\gamma_C \mathcal{I}^3(\Omega),
\end{align}
with $\Omega=(\cos(\theta),\phi)$ the decay angles of the $\rho$ to 2 pions, $\Phi$ is the linear polarization angle with respect to the production plane of the $\rho$-proton final state, $P^\gamma_L$ the degree of linear polarization and $P^\gamma_C$ the degree of circular polarization. Then,

\begin{align} \label{eq:IntensityMoments}
    \mathcal{I}^\alpha = \kappa \sum_{L,M} H^\alpha(LM) Y^m_\ell(\Omega), 
\end{align}

where $Y^m_\ell(\Omega)$ are the Spherical Harmonic functions for angular momentum L and projection M, $H^\alpha(LM)$ the moments of the particular Spherical Harmonic and $\kappa$ a phase space factor.

The moments can be expressed in terms of the spin-density matrix elements (SDMEs) in the reflectivity basis, labelled by their angular momentum $l$ (equal to the particle spin) and spin projection $m$, and the appropriate Clebsh-Gordan coefficients $C$,

\begin{align} \label{eq:H_rho}
    H^\alpha(LM) = \sum_{\ell\ell',mm'}  \left( \frac{2\ell'+1}{2\ell+1} \right) ^{\frac{1}{2}}  C^{\ell 0}_{\ell' 0L0}  C^{\ell m}_{\ell' m'LM} \rho^{\alpha,\ell\ell'}_{mm'},
\end{align}

where the SDMEs are the sum of the two reflectivity components :
\begin{align} \label{eq:reflSDME}
    \rho^{\alpha,\ell\ell'}_{mm'} = \hspace{0.2cm}^{+}\rho^{\alpha,\ell\ell'}_{mm'} \hspace{0.2cm}+\hspace{0.2cm} ^{-}\rho^{\alpha,\ell\ell'}_{mm'},
\end{align}

which are given by,

\begin{align} \label{eq:rho_pws}
    ^{\epsilon}\rho^{0,\ell\ell'}_{mm'} =&
        [\ell]^{(\epsilon)}_{m}[\ell]^{(\epsilon)*}_{m'}
        +(-1)^{m-m'}[\ell]^{(\epsilon)}_{-m}[\ell]^{(\epsilon)*}_{-m'},\notag\\
    ^{\epsilon}\rho^{1,\ell\ell'}_{mm'} =& -\epsilon \left(
        (-1)^{m}[\ell]^{(\epsilon)}_{-m}[\ell]^{(\epsilon)*}_{m'}
        +(-1)^{m'}[\ell]^{(\epsilon)}_{m}[\ell]^{(\epsilon)*}_{-m'} \right), \notag\\
    ^{\epsilon}\rho^{2,\ell\ell'}_{mm'} =&-i\epsilon \left(
        (-1)^{m}[\ell]^{(\epsilon)}_{-m}[\ell]^{(\epsilon)*}_{m'}
        -(-1)^{m'}[\ell]^{(\epsilon)}_{m}[\ell]^{(\epsilon)*}_{-m'} \right),\notag\\
    ^{\epsilon}\rho^{3,\ell\ell'}_{mm'} =&
        [\ell]^{(\epsilon)}_{m}[\ell]^{(\epsilon)*}_{m'}
        -(-1)^{m-m'}[\ell]^{(\epsilon)}_{-m}[\ell]^{(\epsilon)*}_{-m'}.\notag\\
\end{align}
with $[\ell]$ = S, P , ... . For $\rho$ production the P-waves will dominate.

For completeness we give the full expansions of the moments in terms of the S and P waves (note the Clebsch-Gordan factors are given by numerical approximations).

\begin{align*}
     H^{0}(0,0)=&2(^{+}S^{+}S)+2(^{-}S^{-}S)+2(^{+}P_{-1}^{+}P_{-1})+2(^{+}P_{+1}^{+}P_{+1})\\& +2(^{-}P_{-1}^{-}P_{-1})+2(^{-}P_{+1}^{-}P_{+1})+2(^{+}P_{0}^{+}P_{0})+2(^{-}P_{0}^{-}P_{0})\\
    H^{1}(0,0)=&2(^{+}S^{+}S)+-2(^{-}S^{-}S)+-4(^{+}P_{+1}^{+}P_{-1})\cos(^{+}\phi_{P+1}-^{+}\phi_{P-1}))\\& +4(^{-}P_{+1}^{-}P_{-1})\cos(^{-}\phi_{P+1}-^{-}\phi_{P-1}))+2(^{+}P_{0}^{+}P_{0})+-2(^{-}P_{0}^{-}P_{0})\\
    H^{0}(1,0)=&2.3(^{+}S^{+}P_{0})\cos(^{+}\phi_{S}-^{+}\phi_{P0}))\\&+2.3(^{-}S^{-}P_{0})\cos(^{-}\phi_{S}-^{-}\phi_{P0}))\\
    H^{1}(1,0)=&2.3(^{+}S^{+}P_{0})\cos(^{+}\phi_{S}-^{+}\phi_{P0}))\\&-2.3(^{-}S^{-}P_{0})\cos(^{-}\phi_{S}-^{-}\phi_{P0}))\\
    H^{0}(1,1)=&-1.2(^{+}S^{+}P_{-1})\cos(^{+}\phi_{S}-^{+}\phi_{P-1}))+1.2(^{+}S^{+}P_{+1})\cos(^{+}\phi_{S}-^{+}\phi_{P+1}))\\&-1.2(^{-}S^{-}P_{-1})\cos(^{-}\phi_{S}-^{-}\phi_{P-1}))+1.2(^{-}S^{-}P_{+1})\cos(^{-}\phi_{S}-^{-}\phi_{P+1}))\\
    H^{1}(1,1)=&-1.2(^{+}S^{+}P_{-1})\cos(^{+}\phi_{S}-^{+}\phi_{P-1}))+1.2(^{+}S^{+}P_{+1})\cos(^{+}\phi_{S}-^{+}\phi_{P+1}))\\&+1.2(^{-}S^{-}P_{-1})\cos(^{-}\phi_{S}-^{-}\phi_{P-1}))+-1.2(^{-}S^{-}P_{+1})\cos(^{-}\phi_{S}-^{-}\phi_{P+1}))\\
    H^{2}(1,1)=&-1.2(^{+}S^{+}P_{-1})\cos(^{+}\phi_{S}-^{+}\phi_{P-1}))+-1.2(^{+}S^{+}P_{+1})\cos(^{+}\phi_{S}-^{+}\phi_{P+1}))\\&+1.2(^{-}S^{-}P_{-1})\cos(^{-}\phi_{S}-^{-}\phi_{P-1}))+1.2(^{-}S^{-}P_{+1})\cos(^{-}\phi_{S}-^{-}\phi_{P+1}))\\
    H^{3}(1,1)=&1.2(^{+}S^{+}P_{-1})\sin(^{+}\phi_{S}-^{+}\phi_{P-1}))+1.2(^{+}S^{+}P_{+1})\sin(^{+}\phi_{S}-^{+}\phi_{P+1}))\\&+1.2(^{-}S^{-}P_{-1})\sin(^{-}\phi_{S}-^{-}\phi_{P-1}))+1.2(^{-}S^{-}P_{+1})\sin(^{-}\phi_{S}-^{-}\phi_{P+1}))\\
    H^{0}(2,0) =&-0.4(^{+}P_{-1}^{+}P_{-1})+-0.4(^{+}P_{+1}^{+}P_{+1})+-0.4(^{-}P_{-1}^{-}P_{-1})\\&-0.4(^{-}P_{+1}^{-}P_{+1})+0.8(^{+}P_{0}^{+}P_{0})+0.8(^{-}P_{0}^{-}P_{0})\\
    H^{1}(2,0)=&0.8(^{+}P_{+1}^{+}P_{-1})\cos(^{+}\phi_{P+1}-^{+}\phi_{P-1}))-0.8(^{-}P_{+1}^{-}P_{-1})\cos(^{-}\phi_{P+1}-^{-}\phi_{P-1})) \\&+0.8(^{+}P_{0}^{+}P_{0})-0.8(^{-}P_{0}^{-}P_{0})\\
    H^{0}(2,1)=&-0.7(^{+}P_{0}^{+}P_{-1})\cos(^{+}\phi_{P0}-^{+}\phi_{P-1}))+0.7(^{+}P_{0}^{+}P_{+1})\cos(^{+}\phi_{P0}-^{+}\phi_{P+1}))\\&-0.7(^{-}P_{0}^{-}P_{-1})\cos(^{-}\phi_{P0}-^{-}\phi_{P-1}))+0.7(^{-}P_{0}^{-}P_{+1})\cos(^{-}\phi_{P0}-^{-}\phi_{P+1}))\\
    H^{1}(2,1)=&-0.7(^{+}P_{0}^{+}P_{-1})\cos(^{+}\phi_{P0}-^{+}\phi_{P-1}))+0.7(^{+}P_{0}^{+}P_{+1})\cos(^{+}\phi_{P0}-^{+}\phi_{P+1}))\\&+0.7(^{-}P_{0}^{-}P_{-1})\cos(^{-}\phi_{P0}-^{-}\phi_{P-1}))-0.7(^{-}P_{0}^{-}P_{+1})\cos(^{-}\phi_{P0}-^{-}\phi_{P+1}))\\
    H^{2}(2,1)=&-0.7(^{+}P_{0}^{+}P_{-1})\cos(^{+}\phi_{P0}-^{+}\phi_{P-1}))-0.7(^{+}P_{0}^{+}P_{+1})\cos(^{+}\phi_{P0}-^{+}\phi_{P+1}))\\&+0.7(^{-}P_{0}^{-}P_{-1})\cos(^{-}\phi_{P0}-^{-}\phi_{P-1}))+0.7(^{-}P_{0}^{-}P_{+1})\cos(^{-}\phi_{P0}-^{-}\phi_{P+1}))\\
    H^{3}(2,1)=&0.7(^{+}P_{0}^{+}P_{-1})\sin(^{+}\phi_{P0}-^{+}\phi_{P-1}))+0.7(^{+}P_{0}^{+}P_{+1})\sin(^{+}\phi_{P0}-^{+}\phi_{P+1}))\\&+0.7(^{-}P_{0}^{-}P_{-1})\sin(^{-}\phi_{P0}-^{-}\phi_{P-1}))+0.7(^{-}P_{0}^{-}P_{+1})\sin(^{-}\phi_{P0}-^{-}\phi_{P+1}))\\
    H^{0}(2,2)=&-0.98(^{+}P_{+1}^{+}P_{-1})\cos(^{+}\phi_{P+1}-^{+}\phi_{P-1}))\\&-0.98(^{-}P_{+1}^{-}P_{-1})\cos(^{-}\phi_{P+1}-^{-}\phi_{P-1}))\\
    H^{1}(2,2)=&0.49(^{+}P_{-1}^{+}P_{-1})+0.49(^{+}P_{+1}^{+}P_{+1})\\&-0.49(^{-}P_{-1}^{-}P_{-1})-0.49(^{-}P_{+1}^{-}P_{+1})\\
    H^{2}(2,2)=&0.49(^{+}P_{-1}^{+}P_{-1})-0.49(^{+}P_{+1}^{+}P_{+1})\\&-0.49(^{-}P_{-1}^{-}P_{-1})+0.49(^{-}P_{+1}^{-}P_{+1})\\
    H^{3}(2,2)=&0.98(^{+}P_{+1}^{+}P_{-1})\sin(^{+}\phi_{P+1}-^{+}\phi_{P-1}))\\&+0.98(^{-}P_{+1}^{-}P_{-1})\sin(^{-}\phi_{P+1}-^{-}\phi_{P-1}))
\end{align*}

In this case the partial waves are defined by 14 real numbers (2 reflectivities of 1 S wave, 3 P waves ($m=+1,0,-1$), with 2 fixed phases). While we have 18 moments equations, necessary for being over-constrained. Hence we may leave $P^\gamma_L$ and $P^\gamma_C$ as free parameters in our fits.

\begin{figure}
    \centering
    \includegraphics[width=0.8\linewidth]{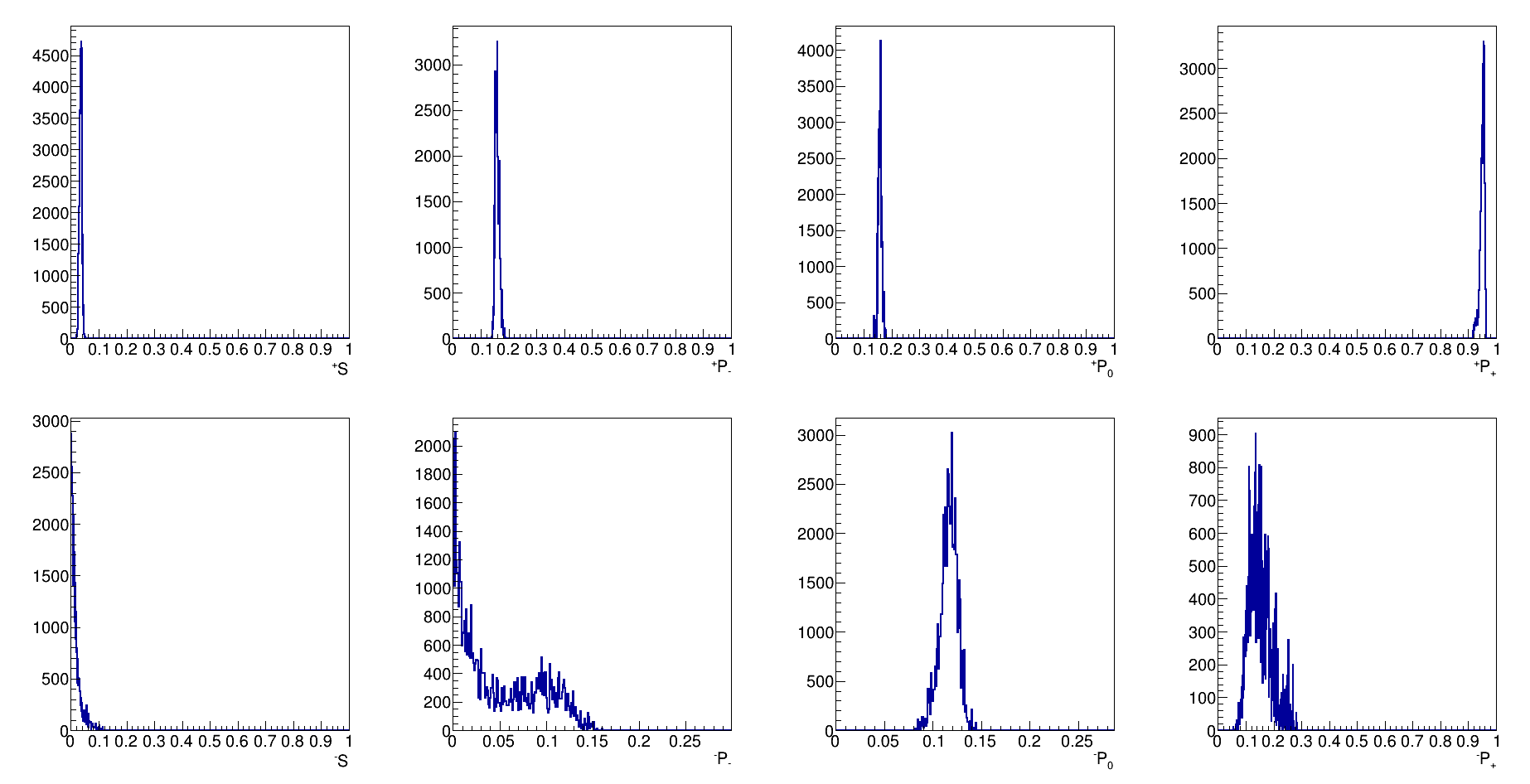}
    \caption{Magnitudes of partial waves extracted in preliminary fits to 2023 data. The top row shows +ve reflectivty amplitudes and bottom -ve. The data distributions show MCMC samples. }
    \label{fig:resultPWAMags}
\end{figure}

\begin{figure}
    \centering
    \includegraphics[width=0.8\linewidth]{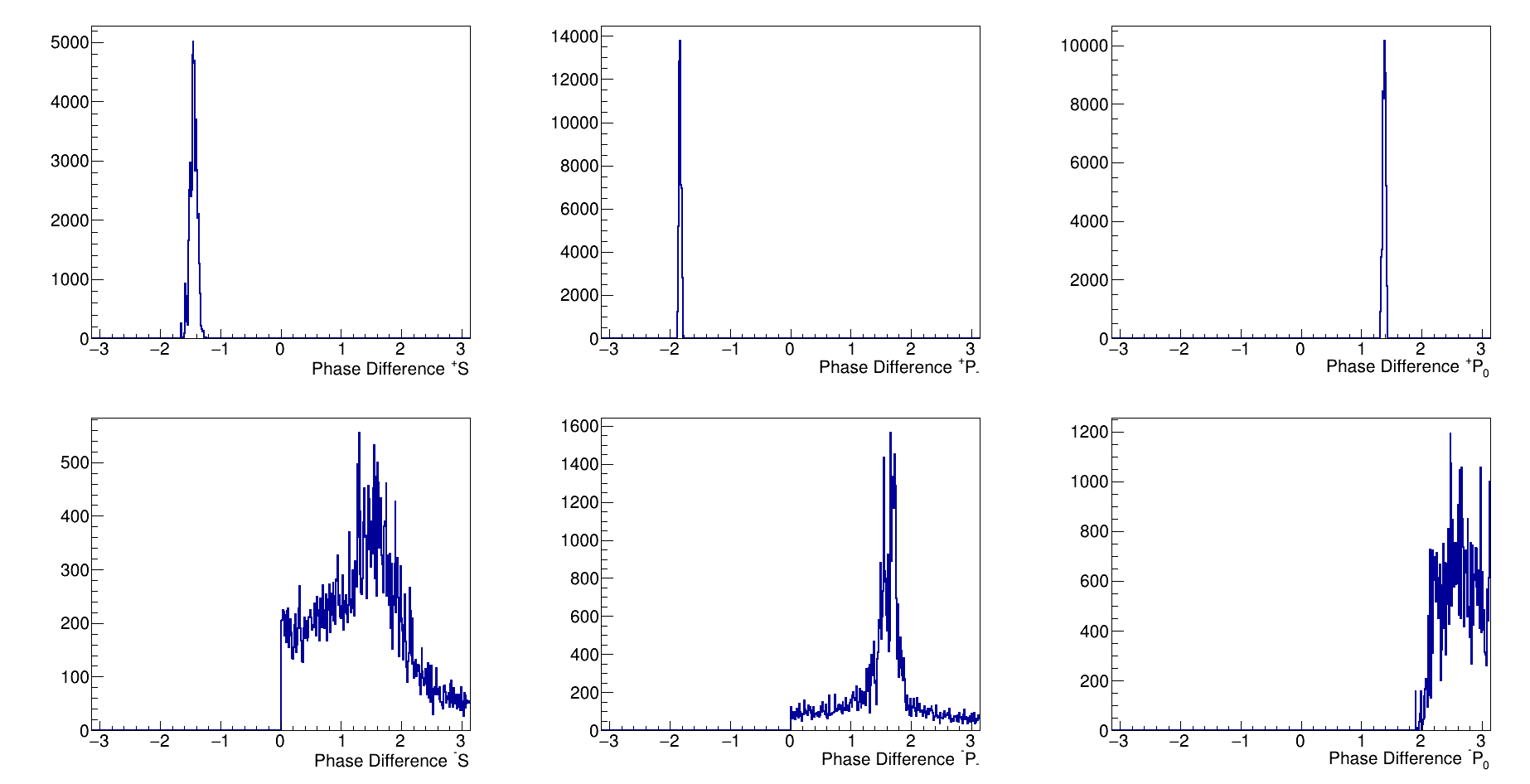}
    \caption{Phases relative to the $P_{+}$ partial waves extracted in preliminary fits to 2023 data. The top row shows +ve reflectivty amplitudes and bottom -ve. The data distributions show MCMC samples. In the case of the -ve reflectivity amplitudes the phases have a large uncertainty due to their small magnitudes.}
    \label{fig:resultPWAPhases}
\end{figure}

The results of our preliminary analysis are shown in Fig.~\ref{fig:resultPLinCirc} for the degrees of polarization. In Figs.~\ref{fig:resultPWAMags} and \ref{fig:resultPWAPhases} we show the corresponding Partial Wave Amplitudes extracted in the same fits. These results are well in line with expectations from the GlueX $\rho$ SDME measurements with a dominant +ve reflectivity and $^{+}P_{+}$ wave.  

\end{document}